\documentclass[11pt,a4paper]{article}
\usepackage{jheppub}
\usepackage[T1]{fontenc}
\usepackage{amssymb}
\usepackage{mathtools}
\usepackage{stackengine}
\usepackage{scalerel}
\usepackage{footmisc}

\usepackage{hyperref}
\usepackage{epsfig}
\usepackage{amsmath,latexsym,amssymb}
\usepackage{graphicx}
\usepackage{braket}
\usepackage{pdfsync}
\usepackage{framed}
\usepackage{mathtools}

\DeclareMathOperator{\Tr}{Tr}
\DeclareMathOperator{\tr}{tr}

\def\be{\begin{equation}}
\def\ee{\end{equation}}
\def\ba{\begin{eqnarray}}
\def\ea{\end{eqnarray}}

\newcommand{\si}{\sigma}

\newcommand{\bz}{\bar{z}}
\newcommand{\bw}{\bar{w}}
\newcommand{\bh}{\bar{h}}

\newcommand{\zbar}{\bar{z}}

\newcommand{\ha}{{1\over 2}}

\def\a{\alpha}

\def\b{\beta}

\def\D{\Delta}
\def\e{\epsilon}

\def\p{\pi}

\def\m{\mu}
\def\n{\nu}
\def\om{\omega}

\def\l{\lambda}

\def\cO{{\cal O}}

\def\cA{{\cal A}}

\newcommand{\comment}[1]{}

\def\fc#1#2{{\frac{#1}{#2}}}
\newcommand{\req}[1]{(\ref{#1})}

 \def\Ac {{\cal A}}

\def\h{\fc{1}{2}}
\def\ov{\overline}

\def\p{\partial}

\newcommand{\eea}{\end{eqnarray}}
\def\lf{\left}
\def\ri{\right}

\def\ra{{\rightarrow}}

\def\Tr{{\rm Tr}}

\author[a]{Wei Fan,}
\author[b,c]{Angelos Fotopoulos,}
\author[d]{Stephan Stieberger}
\author[b] {and Tomasz R. Taylor}

\affiliation[a]{Department of Physics, College of Science, Jiangsu University of Science and Technology, \\ Zhenjiang, 212003, China}
\affiliation[b]{ Department of Physics,	Northeastern University, \\ Boston, MA 02115, USA}
\affiliation[c]{Department of Sciences,	Wentworth Institute of Technology, \\Boston, MA 02115, USA}
\affiliation[d]{Max--Planck--Institut f\"{u}r Physik,	Werner--Heisenberg--Institut, \\80805 M\"unchen, Germany}
\title{\boldmath On Sugawara construction on Celestial Sphere\unboldmath}

\abstract{
	Conformally soft gluons are conserved currents of the Celestial Conformal Field Theory (CCFT) and generate a Kac-Moody algebra. We study celestial amplitudes of Yang-Mills theory, which are Mellin transforms of gluon amplitudes and take the double soft limit of a pair of gluons. In this manner we construct the Sugawara energy-momentum tensor of  the CCFT.
	We verify that conformally soft gauge bosons are Virasoro primaries of the CCFT under the Sugawara energy-momentum tensor. The Sugawara tensor though does not generate the correct conformal transformations for hard states. In Einstein-Yang-Mills (EYM) theory, we consider an alternative construction of the energy-momentum tensor, similar to the double copy construction which relates gauge theory amplitudes with gravity ones. This energy momentum tensor has the correct properties to generate conformal transformations for both soft and hard states. We extend this construction to supertranslations. }


\bigskip
\begin{document}
\maketitle
	
	\section{Introduction}
	\label{sec:intro}

	The null infinity of $D\!=\!4$ asymptotically flat spacetime is the product of a conformal two--sphere (\textit{celestial sphere}) ${\cal C S}^2$ with a null line.  It was realized some time ago, that for asymptotically flat space-times the Poincare group can be extended to the local BMS group. The local BMS algebra, is an infinite dimensional extension of the Poincare algebra. It contains the Virasoro algebra generators, which generate the conformal group, and in addition supertranslations. This is very suggestive of a conformal field theory living on the null infinity of Minkowski space-time.
	
	Indeed, scattering amplitudes in $D\!=\!4$ Minkowski spacetime can be recast, via a Mellin transform, into conformal correlation functions (\textit{celestial amplitudes}) on the celestial sphere~\cite{Pasterski:2016qvg,Pasterski:2017kqt,Pasterski:2017ylz,Schreiber:2017jsr,Cheung:2016iub,Banerjee:2019prz}\footnote{In theories with gravity the Mellin transform from 4d to the ${\cal C S}^2$ is sensitive to UV divergencies. String amplitudes are known for their soft UV properties. Celestial amplitudes for string theories were discussed in \cite{Stieberger:2018edy,Pate:2019mfs}. The role of UV constraints on scattering amplitudes was discussed recently in \cite{Carrasco:2019qwr}.}.  The theory describing the dynamics of celestial amplitudes is expected to be a novel conformal field theory on ${\cal CS}^2$. The celestial amplitudes, correspond to a subset of the correlators on the celestial conformal field theory (CCFT). One of the main motivations, for studying such a theory,   is the proposal that CCFT is a holographic description of 4d physics in Minkowski spacetime~\cite{Cheung:2016iub,strominger_lectures_2017,Pasterski:2019msg}. This originates from early studies of flat holography~\cite{deboer_holographic_2003} and the BMS algebra~\cite{bondi_gravitational_1962,sachs_gravitational_1962,barnich_symmetries_2010}  of asymptotic symmetries. Recently, the study of CCFT and its properties has led to several advances along various aspects of the proposed theory~\cite{Schreiber:2017jsr,Stieberger:2018edy,Kapec:2016jld, fan_soft_2019, Fotopoulos:2019tpe,Pate:2019lpp, Fotopoulos:2019vac,distler_double-soft_2019, Nandan:2019jas,Guevara:2019ypd,Adamo:2019ipt,Laddha:2020kvp,Cardona:2019ywu}, but a lot remains in order to make the CCFT a solid proposal for flat space-time holography.
	
	In CCFT, a particle is described~\cite{Pasterski:2016qvg,Pasterski:2017kqt} by conformal primary wave function. These functions are labeled by the conformal weights $(h,\bar{h})$  and position $(z,\bar{z})$. The coordinates $(z,\bar{z})$ on the celestial sphere are related to the asymptotic direction of  the four-momentum in 4d Minkowski spacetime. The scaling dimension $\Delta$ and spin helicity $J$ can be obtained from the conformal weights via $\Delta = h+ \bar{h}, J=h-\bar{h}$. In CCFT, each particle corresponds to a conformal field operator with ${\rm Re}(\Delta)=1$, i.e.\ $\Delta=1+i\lambda, \lambda\in\mathbb{R}$ \cite{Pasterski:2017kqt}. For massless particles, conformal field operators are Mellin transforms of plane wave functions in the 4d Minkowski spacetime~\cite{Pasterski:2017kqt,Cheung:2016iub}.
	
	At this stage, many details of the CCFT are under investigation. From current studies nontrivial and elegant features have already been discovered, especially its algebraic structure extracted from celestial operator spectra and celestial amplitudes. At the classical level, the Ward identities of conserved currents and energy-momentum tensor have been shown to correspond to soft theorem of gluons and gravitons \cite{strominger_asymptotic_2014,strominger_bms_2014,Cheung:2016iub,Conde:2016rom,Mao:2017tey,Himwich:2019dug,Banerjee:2019tam,Puhm:2019zbl}.
For gluons, conserved currents of the CCFT correspond to the conformal soft limit~\cite{donnay_conformally_2019,Pate:2019mfs,Law:2019glh} of conformal operators, i.e.  $\Delta=1~(\lambda=0)$.\footnote{Any operator with $\D\neq1, (\lambda\neq 0)$ is called a hard operator for the purposes of this work}  A similar picture holds between conformally soft gravitons and the energy-momentum tensor \cite{Cheung:2016iub,Kapec:2016jld}. The studies of the conformal soft limit of celestial amplitudes allowed  making a direct connection between Ward identities of currents on ${\cal CS}^2$ and the  low energy theorems of gluons~\cite{fan_soft_2019} and gravitons  ~\cite{fan_soft_2019,Pate:2019lpp,Fotopoulos:2019tpe}.  Moreover, the collinear singularities of gluon and graviton amplitudes, correspond to the case where two operators of the CCFT approach each other. The study of collinear singularities of celestial amplitudes was used in \cite{fan_soft_2019,Fotopoulos:2019vac} to derive  the operator product expansion (OPE) of celestial operators. Similar work has appeared recently for massive states \cite{Law:2020tsg}.
	
	In this work  we want in particular to elaborate on the energy-momentum tensor $T(z)$  which generates the Virasoro algebra on the ${\cal CS}^2$.	The energy-momentum tensor $T(z)$ is a $\Delta=2$  conformal field operator that can be constructed through a shadow transformation from the $\Delta=0$ operator of the graviton~\cite{Kapec:2016jld,Cheung:2016iub}.  The OPE of this energy-momentum tensor with the conformal field operators of gluons and gravitons was  derived in ~\cite{Fotopoulos:2019tpe,Fotopoulos:2019vac} and indeed it was found that these conformal operators transforms as  Virasoro primary fields. In addition, OPEs of   all the BMS generators, superrotations and supertranslations, were derived~\cite{ Fotopoulos:2019vac}\footnote{A study of representations of the BMS algebra on ${\cal CS}^2$ was initiated in \cite{Banerjee:2020kaa}.}. The study of  these OPEs allowed us to derive the BMS algebra~\cite{Barnich:2017ubf} of asymptotic symmetries.
	%The Virasoro algebra derived has zero central charge (so it is the Witt algebra). We know that the central charge is related to anomaly cancellation of CFT. The CCFT is proposed to be the hologram of flat spacetime, so it should be a consistent physics theory, which should have zero central charge such that all anomaly cancel out. On the one hand, this zero central charge~\cite{Fotopoulos:2019vac} is computed in  tree-level amplitudes, and the quantum corrections are still unknow.
	
	An alternative proposal for the energy-momentum tensor of a pure gluon theory appeared in \cite{He:2015zea}.  It was shown that positive helicity soft gluons correspond to holomorphic conserved currents on the ${\cal CS}^2$ which generate a $D\!=\!2$ Kac-Moody algebra.  In standard CFT, in the presence of a Kac-Moody algebra, we can use the Sugawara construction \cite{Sugawara:1967rw} to build the energy-momentum tensor. It is natural to ask ourselves if it is possible to extend this construction on the CCFT. Some initial attempts in this direction using double soft theorems of gluons appeared in \cite{McLoughlin:2016uwa,Nande:2017dba}. These results showed several interesting features and short-comings of the Sugawara construction. In the current paper we will approach this problem from the point of view of celestial amplitudes. We will construct the Sugawara tensor from the double conformal soft limit of gluons. Furthermore, we will derive its OPE with conformal operators of gluons and discuss its properties depending on whether the operators are soft or hard. We will conclude that the Sugawara energy-momentum tensor can only capture the dynamics of the soft sector of the theory confirming earlier observations in \cite{Cheung:2016iub}. In the setup of an Einstein-Yang-Mills (EYM) theory, we will present an alternative approach, based on a pair of gluon conformal operators. This is reminiscent of the double copy or Kawai--Lewellen--Tye (KLT)  construction of gravity amplitudes from gauge theory amplitudes. This energy momentum tensor captures the dynamics of soft and hard operators alike. We will see that we can extend this construction to include the supertranslation generators as well.
	
	This article is organized as follows. In Section~\ref{sec:Prelim}, we review the notation, useful formulas of the CCFT and the Mellin transform of 4d gluon amplitudes which generates celestial amplitudes. We also review the Kac-Moody algebra on ${\cal CS}^2$ and the proposed Sugawara energy-momentum tensor based on conformally soft gluon operators. In Sections~\ref{sec:Sugawara_ope} and \ref{sec:coincide_result}, we compute the celestial amplitudes with a Sugawara energy-momentum  tensor insertion. We discuss in details the case of $SU(N)$ and general gauge groups studying the Mellin transform of MHV gluon amplitudes \cite{Parke:1986gb} under the double conformal soft limit. We derive the OPE of the energy momentum tensor with conserved currents. Our results agree with standard Kac-Moody current algebra expectations. In section \ref{sec:gengroup} we re-derive the OPEs of the Sugawara tensor applying the OPE between currents and then taking the conformal soft limits. Our results are consistent. Finally, in section \ref{sec:shadow} we derive an energy momentum tensor which is inspired by the double copy or KLT construction of gravity amplitudes from gauge amplitudes. This energy-momentum tensor captures the conformal properties of both hard and soft operators.

	\section{Remarks on gluon operator products and Sugawara construction}
	\label{sec:Prelim}

	The $D=4$ momentum of a massless particle is parametrized by coordinates $(z,\bar{z})$ of the celestial sphere as
	\begin{equation}
	\label{eq:pz}
	p^{\mu}=\epsilon \omega q^{\mu}, \quad q^{\mu}=\frac{1}{2}\left(1+|z|^{2}, z+\bar{z},-i(z-\bar{z}), 1-|z|^{2}\right),
	\end{equation}
	with $\omega$  the light-cone energy and $\epsilon=\pm$ indicating outgoing/incoming particles. The  asymptotic direction along which the particle propagates is given by the null vector $q^\mu({z,\bar{z}})$. This vector is parametrised by the coordinates $(z,\bar{z})$ on the celestial sphere. In working on the celestial sphere we will need to transform plane wave solutions to conformal primary wave functions~\cite{Pasterski:2017kqt}.  For gluons, a conformal primary wave function with conformal dimension $(h,\bar{h})$ is given by the following expression:
	\begin{equation}
	\label{eq:primary_wave}
	A_{\mu}^{\Delta, J}=g(\Delta) V_{\mu J}^{\Delta}+\text {pure gauge term},
	\end{equation}
	where $g(\Delta)=\tfrac{(\Delta-1)}{\Gamma(\Delta+1)}$ is the normalization constant,  $\Delta=h+\bar{h}$ is the scaling dimension and $J=h-\bar{h}=\pm 1$ is the spin helicity. The function   $V_{\mu J}^{\Delta}$ is the Mellin transform of the 4d plane wave function
	\begin{equation}
	\label{eq:mellin}
	V_{\mu}^{\Delta, J}\left(X^{\mu}, z, \bar{z}\right) \equiv \partial_{J} q_{\mu} \int_{0}^{\infty} d \omega\ \omega^{\Delta-1} e^{\mp i \omega q \cdot X-\varepsilon \omega}\ , \qquad J=\pm 1\ ,
	\end{equation}
	where $\partial_J=\partial_z$ for $J=+1$ and $\partial_J=\partial_{\bz}$ for $J=-1$. The polarization vectors are $\partial_{z} q^{\mu}=\epsilon_{+}^{\mu}(p)$ and $\partial_{\bar{z}} q^{\mu}=\epsilon_{-}^{\mu}(p)$. The scaling dimensions are %belongs to the  principle continuous series,
	$\Delta=1+i\lambda, \lambda\in\mathbb{R}$,
	%of irreducible unitary representations of the Lorentz group
	~\cite{Pasterski:2017kqt}.
	In a similar manner for gravitons the conformal primary wave function is
	
\be\label{confprimMellin2}
H^{\D,\ell}_{\mu \n} (X^\mu, z, \bz)\equiv \partial_J q_\mu \partial_J q_\nu \int _0^\infty d\omega  \ \omega^{\D-1} e^{\mp i \omega q \cdot  X -\e \omega}\ ,\qquad \ell=\pm 2\ .
\ee
where $J=+1$ for $\ell=+2$ and $J=-1$ for $\ell=-2$.
The conformal (quasi-primary) wave functions can be written as
\be\label{confprimexpr3}
G^{\D,\ell}_{\m \n }=
f(\D)H^{\D,\ell}_{\mu \n} +\makebox{diff}
\ee
with the normalization constant $ f(\D)=
{1\over 2}{\D(\D-1) \over \Gamma(\D+2)}$.
The presence of these normalization factors makes it clear that, as mentioned in the introduction, fields with spin 1 become pure gauge when $\D=1$ while fields with spin 2 become pure diffeomorphisms for $\D=0,1$.

	In this work we will study $D=4$ tree--level gluon amplitudes\footnote{Here we use capital $A$ for gluon amplitudes, with $A_n$ representing the full amplitude and $A^\sigma$ representing the partial amplitude. We choose to use the calligraphic $\mathcal{A}$ for Mellin/celestial amplitude. This is different convention from reference~\cite{fan_soft_2019}, where the calligraphic $\mathcal{M}$ is used for gluon partial amplitudes. } $A_{n}$ and their celestial sphere representation $\Ac_n$.
For a generic gauge group $G$ the $D=4$ gluon scattering amplitudes can be expressed as a sum over  partial subamplitudes as follows\footnote{In the amplitudes community, color generators $T^a$ differ from the mathematics definition $t^a$ by a  factor of $\sqrt 2$ absorbed into each generator, i.e. $T^a=\sqrt2 t^a$. As a consequence, for the Lie algebra $g$ the commutation relation $[t^a,t^b]=i f^{abc}t^c$ implies $[T^a,T^b]=i \tilde f^{abc}T^c$ with the following dictionary  for the structure constants $\tilde f^{abc}=\sqrt 2 f^{abc}$. See also footnote \ref{grouptheory} for further details.}
	\begin{equation}
	\label{eq:4d_amp}
	A_{n}(\{\om_i, q_i,J_i\}) = \sum_{\sigma \in S_{n-1}} \Tr\left(T^{a_1}T^{a_{\sigma(2)}}\ldots T^{a_{\sigma(n)}}\right) A^{\sigma}_{J_1J_2\ldots J_{n}}\left(\{\omega_{i}, z_{i}, \bar{z}_{i}\}\right)\ ,
	\end{equation}
with $T^a$ gauge generators in the fundamental representation of the gauge group $G$, the spin helicities denoted by $J_i=\pm1,i=1,2,\ldots,n$ and $A^{\sigma}_{J_1J_2\ldots J_{n}}$ is the partial subamplitude for a given permutation $\sigma$ expressed in celestial coordinates $\{\omega_{i}, z_{i}, \bar{z}_{i}\}$.	
The CCFT amplitudes are identified with the space--time amplitudes transformed from the plane wave basis  into the conformal basis  (\ref{eq:primary_wave},\ref{confprimexpr3}) by using properly normalized Mellin transform\cite{Pasterski:2016qvg,Pasterski:2017ylz,Schreiber:2017jsr,
Stieberger:2018edy}. Concretely, the gluon partial subamplitudes $A^{\sigma}_{J_1J_2\ldots J_{n}}$  give rise to the celestial gluon amplitude:
\begin{align}
	\label{eq:mellin_partial_amp}
	\mathcal{A}^{\sigma}_{J_{1} \ldots J_{n}}\left\{(\Delta_{i}, z_{i}, \bar{z}_{i}\}\right)&=\left(\prod_{i=1}^{n} g\left(\Delta_{i}\right) \int_0^\infty d \omega_{i}\ \omega_{i}^{\Delta_1-1}\right) A^{\sigma}_{J_{1} \ldots J_{n}}\left(\{\omega_{i}, z_{i}, \bar{z}_{i}\}\right)\delta^{4}\left(\sum_{i=1}^{n}\epsilon_{i}\omega_{i} q_{i}\right)\ .
	\end{align}
 In Eq.(\ref{eq:mellin_partial_amp}) $\epsilon_i= +1$ or $-1$ depending whether the particles are incoming or outgoing, respectively. 	
The full CCFT correlator is identified with the S--matrix element transformed from the plane wave basis  into conformal basis:
	\begin{equation}
	\label{eq:celestial_amp}
	\left\langle\mathcal{O}_{\Delta_{1} J_{1}}^{a_{1}} \mathcal{O}_{\Delta_{2} J_{2}}^{a_{2}}\ldots \mathcal{O}_{\Delta_{n} J_{n}}^{a_{n}}\right\rangle =\sum_{\sigma \in S_{n-1}} \mathcal{A}_{J_{1} J_{2} \cdots J_{n}}^{\sigma} \operatorname{Tr}\left(T^{a_{1}} T^{a_{\sigma(2)}} \ldots T^{a_{\sigma(n)}}\right):=
	\Ac_n\left(\{\Delta_{i}, z_{i}, \bar{z}_{i},J_i\}\ri),
	\end{equation}
	where $\mathcal{O}_{\Delta_{i} J_{i}}^{a_{i}}$ is the conformal field operator which corresponds  to a gluon conformal primary wave function (\ref{eq:primary_wave}).

%%%%  	\ba\label{corrdef}
%%%% \Big\langle \prod_{i=1}^n\cO_{\D_i,J_i}(z_i,\zbar_i)\Big\rangle&=&\Big(  \prod_{n=1}^n %%%%  g_i(\D_i) \int d \omega_i  \ \omega_i^{\D_i-1} \Big)  \d^{(4)}\big(\sum_{i=1}^n  %%%% \epsilon_i\om_i q_i\big)\nonumber\\[1mm] \label{cftcor}
%%%%  &&~~~~~\times
%%%%  \cM_{J_1\dots J_n}(\omega_i, z_i, \bz_i)\ .
%%%%   \ea

	On the celestial sphere, the limit $z\to w$ of coinciding positions for two operators,  corresponds to $q^\mu(z)\to q^\mu(w)$ for the 4d gluon particles. This limits corresponds to the collinear momentum limit $p^\mu(z)\parallel p^\mu(w)$. It is well known that gauge and gravity amplitudes have collinear singularities and based on the discussion above, they give rise to the OPE singularities of the holographic CCFT. In \cite{fan_soft_2019, Fotopoulos:2019vac} (see also \cite{Pate:2019lpp}), using collinear limits of the 4d gluon amplitude,  it was shown that the CCFT has the following OPEs for gluon conformal primaries
\be\label{opepp}
\cO^{a}_{\D_1, +}(z,\zbar)\,\cO^{b}_{\D_2,+}(w,\bar w)
=\frac{C_{(+,+)+}(\D_1,\D_2)}{z-w} f^{abc}\cO^{c}_{(\D_1+\D_2-1),+}(w,\bar w)+\makebox{regular}\, ,
\ee
with
\be C_{(+,+)+}(\D_1,\D_2)=1-\frac{(\D_1-1)(\D_2-1)}{\D_1\D_2}\label{ope1}\ ,
\ee
and
\ba
\cO^{a}_{\D_1 ,+}(z,\zbar)&&\!\!\!\!\cO^{b}_{\D_2,-}(w,\bar w)=\nonumber\\[2.5mm]
&&~~\frac{C_{(+,-)-}(\D_1,\D_2)}{z-w} f^{abc}\cO^{c}_{(\D_1+\D_2-1),-}(w,\bar w)~~~~\label{opemp}\\[1.1mm]
&&+\,\frac{C_{(+-)+}(\D_1,\D_2)}{\bz-\bw} f^{abc}\cO^{c}_{(\D_1+\D_2-1),+}(w,\bar w)\nonumber\\[.8mm]&& +\;C_{(+-)--}(\D_1,\D_2)\frac{\bar z-\bar w}{z-w}\,\delta^{ab}\,{\cal O}_{(\D_1+\D_2),-2}(w,\bar w)\nonumber\\[2mm]
&&+\; C_{(+-)++}(\D_1,\D_2)\frac{z-w}{\bar z-\bar w}\,\delta^{ab}\,{\cal O}_{(\D_1+\D_2),+2}(w,\bar w)\, +\,\makebox{regular}\, ,\nonumber
\ea
with:
\ba C_{(+,-)-}(\D_1,\D_2)&=&\frac{\D_2-1}{\D_1(\D_1
+\D_2-2)}\label{ope3}\ ,\nonumber \\[1mm]
C_{(+,-)+}(\D_1,\D_2)&=&\frac{\D_1-1}{\D_2(\D_1
+\D_2-2)}\label{ope4}\ ,\nonumber
\\[2mm]
C_{(+-)--}(\D_1,\D_2)&=& - \frac{2(\D_2-1)(\D_2+1)(\D_1-1)}{\D_1(\D_1+\D_2)(\D_1+\D_2-1)}          \ , \\[2mm]
C_{(+-)++}(\D_1,\D_2) &=& - \frac{2(\D_1-1)(\D_1+1)(\D_2-1)}{\D_2(\D_1+\D_2)(\D_1+\D_2-1)}          \ .\nonumber
\ea
The subleading terms \footnote{In our units, the gravitational and gauge coupling constants $\kappa=2$ and $g_{\rm YM}=1$, respectively.} of the mixed helicity OPE are associated with corrections of the EYM theory and won't be important in the pure YM case. Nevertheless, we present these terms here since they will be important in the shadow construction of the energy momentum tensor of section~\ref{sec:shadow}.

In this work we will explore further the properties of a class of correlators, which involve the conformally soft gluon operators $\l \to 0$ ~\cite{Donnay:2018neh,fan_soft_2019}. On the CCFT side these lead to $\D=1$ conserved currents:
	\begin{equation}
	\label{eq:conserved_current}
	j^{a}(z)=\mathcal{O}_{\Delta=1,J=+}^{a}(z, \bar{z}), \quad \bar{j}^{a}(\bar{z})=\mathcal{O}_{\Delta=1,J=-}^{a}(z, \bar{z}).
	\end{equation}
	The conserved currents suggest an emerging infinite dimensional symmetry algebra, commonly known as Kac-Moody current algebra.
The soft limit $\D_1 \to1$  (\ref{opepp}) leads to the following OPE on gluon conformal primaries:
 \be\label{eq:JOope}
 j^a(z) O^b_{\D,+}(w) \sim {f^{a b c}\ O^c_{\D,+}(w) \over z-w}\ .
 \ee
 In the case of same helicity gluons the consecutive soft limit is equivalent to the double soft limit.
 Taking the consecutive soft  limit $\D_1,\D_2\to 1$ in (\ref{opepp}), we are led to the holomorphic current algebra:
  \be\label{eq:JJope}
 j^a(z) j^b(w) \sim {f^{a b c}\ j^c(w) \over z-w}\ .
 \ee
On the other hand, the soft limit $\D_1$ on the mixed OPE (\ref{opemp}) leads to the following result
\be\label{eq:JOmope}
j^a(z) O^{b}_{\D,-}(w)\sim {f^{a b c}\ O^c_{\D,-}(w) \over z-w}
\ee
and similar results for $\bar{j}^a(\bz)$.
Now taking the consecutive double soft limit $\D_1,\D_2 \to 1$ of the mixed helicity gluon OPE we see that the result depends on the order of limits. Specifically taking the soft limit of the positive helicity gluon always first we get
\be\label{eq:jbjope}
j^a(z) \bar{j}^{b}(\bw)\sim {f^{a b c}\  \bar{j}^c(\bw) \over z-w} \ ,
\ee
and taking the negative one first followed by the positive we get:
\be\label{eq:bjjope}
\bar{j}^a(\bz) j^{b}(w)\sim {f^{a b c}\ j^c(w) \over\bz-\bw} \ .
\ee
The order of the soft limits is crucial
%and we will adopt in the OPE that the order of soft limits follows the order of taking soft first the operator on the left and then on the right. This is only important
when we have opposite helicity states or equivalently opposite spin operators.

The relations above imply that the antiholomorphic currents $\bar{j}^a(\bz)$ transform in the adjoint representation of the Kac-Moody symmetry generated by the holomorphic currents $j^a(z)$ and vice-versa. As explained in more details in \cite{He:2015zea} a symmetric limit which realizes  both the holomorphic and antiholomorphic Kac-Moody non-Abelian algebras is not possible. This seems to be related to 3d Chern-Simons theory on a manifold with boundary. The theory naively has two gauge connections $A_z $ and $A_{\bz}$, which generate Kac-Moody symmetries, but in
the non-Abelian case, boundary conditions eliminate one of them leaving only one copy. In \cite{Cheung:2016iub} this idea was further explored. The 4d Minkowski spacetime is written as a foliation of $AdS_3$ slices. There,  it was demonstrated that indeed the soft sector of the theory leads to a CS theory on the $AdS_3$ slices. For the non-Abelian case boundary conditions allow either positive or negative helicity gluons. The $AdS_3/CFT_2$ correspondence implies only a single copy of a Kac-Moody algebra for the soft gluon sector.  We conclude that in the CCFT we need to consider correlators where either the positive or the negative helicity gluons are conformally soft, but not both. When discussing the energy momentum tensor, we chose to study the realisation of the holomorphic Kac-Moody algebra generated by $j^a(z)$ in (\ref{eq:conserved_current}).

It is known that for two dimensional $CFT_2$ the energy momentum tensor for affine current algebras is given by the Sugawara construction. 	As mentioned before we expect only one copy of the Kac-Moody algebra and therefore only one Sugawara energy momentum tensor. The soft sector of positive helicity gauge bosons forms a sub-$CFT_2$ of the full CCFT. Hard particles are sources of soft radiation\footnote{As explained in \cite{He:2015zea, Cheung:2016iub} hard sources can be described by Wilson lines in the 4d-gauge theory along the spirit of jet physics \cite{Feige:2014wja}. We will discuss this further in section \ref{sec:comments}.}. On the CCFT  side, correlation functions factorize into a hard and a soft part. The soft part is  expected to be described by a current algebra and its conformal properties encoded in the Ward identities of the Sugawara energy momentum tensor.  In this paper we will construct the Sugawara energy momentum tensor using the double conformal soft limit  of celestial amplitudes like~\eqref{eq:celestial_amp}.  We will consider gluon amplitudes and  study the limit where the Sugawara tensor becomes collinear with conformally soft positive helicity gluons, the holomorphic current algebra currents $j^a(z)$.

The Sugawara construction \cite{Sugawara:1967rw,DiFrancesco:1997nk} gives an expression of the energy momentum tensor in terms of gauge currents
\begin{equation}\label{eq:sugawara_0}
T^S(w)=\fc{1}{2k+C_2}\ \sum_a :J^a(w)J^a(w):\ ,
\end{equation}
where $k$ is the level of the affine current algebra and the quadratic Casimir\footnote{As a consequence we have $\tilde C_2 =2C_2$, with $\tilde C_2$ referring to the structure constants $\tilde f^{abc}$ of the generators $T^a$, i.e. $\tilde C_2=\delta^{ab} \tilde f^{acd}\tilde f^{bcd}$ \label{grouptheory}.} of the adjoint representation is
$C_2=\delta^{ab} f^{acd}f^{bcd}$, which is twice the dual Coxeter number $h(g)$, i.e. $C_2=2h(g)$. Usually, for free fields the normal ordering is achieved by subtracting
the corresponding two--point correlator
\be\label{eq:sugawaranormal}
T^S(w_1)=\fc{1}{2k+C_2}\ \lim\limits_{w_2\rightarrow w_1} \left\{\sum_a J^a(w_1)J^a(w_2)-\frac{k\dim g}{(w_1-w_2)^2}\right\}
\ee
where $\dim g=\sum_a\delta_{aa}$ is the dimension of the underlying gauge group \cite{Polchinski:1998rr}. We assume roots of length--squared two. A more general normal ordering can explicitly be imposed by the
following contour integral:
\begin{equation}\label{eq:Sugawara_1}
T^S(w_1)=\fc{1}{2k+C_2}\ \fc{1}{2\pi i}\oint_{w_1} \fc{dw_2}{w_2-w_1}\ \sum_a J^a(w_2)J^a(w_1)\ .
\end{equation}
This more general definition of normal ordering takes into account all possible singular terms and is appropriate in case of fields which
are not free  \cite{DiFrancesco:1997nk}.

In the following sections we will construct the Sugawara energy momentum tensor first implementing the normal ordering prescription  (\ref{eq:sugawaranormal})
	\begin{equation}
	\label{eq:sugawara_def}
	T^S(z) =\gamma\sum_{a} j^{a}(z) j^{a}(z)=\gamma \lim\limits_{\Delta,\Delta'\to 1} \lim\limits_{z' \to z} \sum_a \mathcal{O}^a_{\Delta, +}(z, \bar{z})\mathcal{O}^a_{\Delta', +} (z', \bar{z}')\ ,
	\end{equation}
	where $\gamma$ is a normalization constant depending on the details of the Yang-Mills theory. For simplicity, we will ignore this normalization constant and determine its value in the end of the computation. We expect that this value will help determine a discrepancy regarding the level $k$ observed in \cite{Cheung:2016iub}. The relation above is to be considered always as an insertion in a celestial CFT correlator. We will demonstrate that in the case of celestial correlators of MHV amplitudes, (\ref{eq:sugawara_def}) is indeed the energy momentum tensor for the sub-CFT of currents $j^a(z)$. The expected OPE of $T^S(z)$ with a conserved current $j^a(w)$ should be
	\begin{equation}
	\label{eq:tj-ope}
	T^S(z)j^a(w) = \frac{1}{(z-w)^{2}} j^a(w)  +\frac{1}{z-w} \partial_{w} j^a(w) +\ldots\ .
	\end{equation}
	%We further on expect that the OPE with the anitholomorphic currents $\bar{j}^a(\bw)$ vanishes.
	%To prove this result in the celestial amplitude~\eqref{eq:celestial_amp}, we  choose $j(w)$ as $j(z_{j}), j=3,4,\ldots,n-1$, and derive the  $T^S(z)j(z_{j})$-OPE, which will be shown in the next section.
	
	The collinear limit of the Sugawara tensor with negative and positive helicity hard states will also be discussed. We will see that conformal invariance of the full CCFT including the hard sources, necessitates additional contributions to the energy momentum tensor beyond the Sugawara construction. Our discussion in section \ref{sec:shadow} extends our construction to a double copy  (or KLT) type energy momentum tensor, where the conformally soft graviton in \cite{Fotopoulos:2019tpe} is described as a pair of conformally soft gluons. This provides an alternative definition of the energy momentum tensor which includes both soft and hard modes on equal footing. It is nevertheless distinct to the Sugawara tensor, since it does not include a bilinear of the dimension one currents $j^a(z)$.

	\section{Gluon amplitudes, gauge current insertion and operator products}
	\label{sec:Sugawara_ope}
	
In this Section we shall discuss gluon amplitudes \req{eq:4d_amp} with insertion of a pair of
gauge currents \req{eq:sugawara_def}.
From the CCFT theory point of view the Sugawara construction (\ref{eq:sugawara_def}) corresponds  to performing the double conformal soft limit of two positive helicity gluons taken to be collinear at the same time.
In order to study the OPEs of this tensor with primaries we start at the $D=4$ tree--level $n+2$--point gluon MHV amplitude $A_{n+2}(\{\om_i, q_i,J_i\})$.
We shall construct the  Sugawara energy--momentum $T^S(z)$ and derive its OPE with conserved currents $j(z)$ in the celestial amplitude~\eqref{eq:celestial_amp}. In the latter we use the conformal primary operators $\mathcal{O}^{a_{n+1}}_{\Delta_{n+1}+}$ and $\mathcal{O}^{a_{n+2}}_{\Delta_{n+2}+}$ of the last two gluons to construct $T^S(z)$. Obviously, the result does not depend on this choice.  Their color indices will be contracted $a_{n+2}=a_{n+1}=a$ and their positions will approach each other by taking the limit $z_{n+1},z_{n+2}\to z$.
In the next step, in section \ref{sec:coincide_result} we shall take the conformal soft limit $\Delta_{n+1}, \Delta_{n+2}\to 1$ to get $T^S(z)\sim j^a(z)j^a(z)$. The OPE with the primaries $j(z)$ is extracted  by taking the coinciding  limit $z \to z_j, j=3,4,\ldots,n$ of $T^S(z)$ with  primaries $\mathcal{O}^{a_j}_{\Delta_j+}$.
	%As commented in~\cite{Cheung:2016iub}, the Sugawara energy-momentum tensor contains the soft sector of the theory, which corresponds to the conformal soft limit~\cite{fan_soft_2019}.
	Finally, since we are interested only in the soft sector, at the end we will take the conformal soft limit $\Delta_j\to 1$ to get the OPE $T^S(z)j^{a_j}(z_j)$.

In order to study the OPEs with primaries we shall focus on the MHV case. Hence, we will restrict \req{eq:4d_amp} to the $D=4$ tree--level $n+2$--point gluon MHV amplitude \cite{Parke:1986gb} $A_{n+2}(\{\om_i, q_i,J_i\})$ with  spin helicities  $J_1=J_2=-1,J_i=+1,i=3,4,\ldots,n+2$ and their corresponding partial subamplitudes $A^{\sigma}_{J_1J_2\ldots J_{n+2}}$ in  celestial sphere representation 	
	\begin{align}
	\label{eq:partial_amp}
	A^{\sigma}_{J_1J_2\ldots J_{n+2}}\left(\{\omega_{i}, z_{i}, \bar{z}_{i}\}\right) &= \frac{\langle12\rangle^4}{\langle1\sigma(2)\rangle\langle\sigma(2)\sigma(3)\rangle\ldots\langle\sigma(n+2)1\rangle} \nonumber\\
	&= \frac{\omega_{1} \omega_{2}}{\omega_{3} \omega_{4} \ldots \omega_{n+2}} \frac{z_{12}^{4}}{z_{1\sigma(2)} z_{\sigma(2)\sigma(3)}\ldots z_{\sigma(n+2)1}}\ ,
	\end{align}
	with $z_{jk}=z_j-z_k$.

To summarize, in the following two subsections we shall demonstrate the following relation
\begin{align}
\lim_{z_{n+1}\ra z_j} A_{n+2}(\{g_{n+2}^+,g_1,&\ldots,g_n,g_{n+1}^+\})
=-\fc{\tilde C_2(G)}{\omega_{n+1}\omega_{n+2}}\nonumber\\
&\times\lf(\fc{1}{(z_{n+1}-z_{j})^2}+
\fc{\tilde\partial_{z_j}}{z_{n+1}-z_{j}}\ri) A_{n}(\{g_1,\ldots,g_{n}\}),\ j=1,\ldots,n\ ,\label{OPEn}
\end{align}
with the full $n$ gluon amplitude $A_{n}(\{g_1,\ldots,g_{n}\})$ and the quadratic Casimir
$$\tilde C_2(G)=2 C_2(G).$$
The relation \req{OPEn} holds
for any gauge group $G$. Above we have introduced  the derivative
\be\label{tderi}
\tilde\partial_{z_j}=\fc{\p}{\p z_j} -4\ \fc{\delta^{j,j_1}-\delta^{j,j_2}}{z_{j_1j_2}}\ ,
\ee
which singles out
the two gluons $j_1,j_2$ with negative helicity. Of course, for \req{eq:partial_amp} we have: $j_1=1,j_2=2$.
For $j\neq j_1,j_2$ we get $\tilde\partial_{z_j}=\partial_{z_j}$ and the relation \req{OPEn} takes the form of \req{eq:mellin_0} further used in section \ref{sec:coincide_result}.

\subsection[Gluon amplitudes and operator products for $SU(N)$]{Gluon amplitudes and operator products for $\mathbf{SU(N)}$}

Let us first discuss\footnote{A similar construction restricted to $SU(N)$ gauge group and considering double soft limit of gluons appears in \cite{McLoughlin:2016uwa}. However, in the latter reference the Mellin representation, which will be determined in section
	\ref{sec:coincide_result}  is not addressed. Moreover, our analysis which is based on the CCFT formulation, will be extended to arbitrary gauge groups in subsection \ref{sec:colorsums} and section \ref{sec:gengroup}, respectively.} the gauge group $G=SU(N)$.
For simplicity, we also include the photon and choose the gauge group\footnote{The final result for $SU(N)$ is the same as $U(N)$. We will explain this point later.} to be $U(N)$. The fundamental representation of  $U(N)$ satisfies the following useful relations:
	\begin{equation}
	\label{eq:group_formula}
	\left(T^{a}\right)_{k}^{j}\left(T^{a}\right)_{l}^{s}=\delta_{l}^{j} \delta_{k}^{s}, \quad
	\left[T^{a}, T^{b}\right]=i\sqrt{2} f^{a b c} T^{c}\equiv i\tilde f^{abc}T^c.
	\end{equation}

In the following computation, we firstly analyze the $D=4$ MHV amplitudes and then perform the Mellin transform in section \ref{sec:coincide_result}. From  the partial amplitude~\eqref{eq:partial_amp}, various poles of the OPE obviously come from the denominators $z_{n+1,n+2}$, $z_{n+2,j}$ and $z_{n+1,j}$. The poles $z_{n+1,n+2}$ are to be subtracted under the normal ordering of $T^S(z)$ (\ref{eq:sugawaranormal}). We will show though that since our tree amplitudes imply that the level $k$ of the Kac-Moody is zero, such a subtraction wont be necessary.  All such $z_{n+1,n+2}$  poles will drop automatically.
	
	 Among all the permutations, the double poles arise in the following 6 kinds of ordering
	\begin{equation}
	\label{eq:pole_double_ordering}
	\begin{array}{ccc}
	A(\ldots,j,n+1,n+2,\ldots)&A(\ldots,n+1,j,n+2,\ldots)&A(\ldots,n+1,n+2,j,\ldots),\\
	A(\ldots,j,n+2,n+1,\ldots)&A(\ldots,n+2,j,n+1,\ldots)&A(\ldots,n+2,n+1,j,\ldots),
	\end{array}
	\end{equation}
	while the single poles arise in the following 12 possible orderings
	\begin{equation}
	\label{eq:pole_single_ordering}
	\begin{array}{ccc}
	A(\ldots,j,n+1,\ldots,n+2,\ldots)&A(\ldots,n+1,\ldots,j,n+2,\ldots)&A(\ldots,n+1,n+2,\ldots,j,\ldots) \nonumber\\
	A(\ldots,n+1,j,\ldots,n+2,\ldots)&A(\ldots,n+1,\ldots,n+2,j,\ldots)&A(\ldots,n+2,n+1,\ldots,j\ldots) \nonumber\\
	A(\ldots,j,\ldots,n+2,n+1,\ldots)&A(\ldots,n+2,j,\ldots,n+1,\ldots)&A(\ldots,n+2,\ldots,n+1,j\ldots) \nonumber\\
	A(\ldots,j,\ldots,n+1,n+2,\ldots)&A(\ldots,j,n+2,\ldots,n+1,\ldots)&A(\ldots,n+2,\ldots,j,n+1\ldots).
	\end{array}
	\end{equation}
	
\ \\
\noindent\underline{The double poles:}\\	
	\label{sec:sub_double}
	
	\noindent
	Now let's analyze the terms that contribute to double poles.  Consider the two kinds of ordering $(\sigma(i),j,n+1,n+2,\sigma(i+1))$ and $(\sigma(i),j,n+2,n+1,\sigma(i+1))$. Using the formula~\eqref{eq:group_formula}, it is straightforward to get the contraction of the  color indices for $U(N)$
	\begin{align}
	\label{eq:color_1}
	%&\Tr(\ldots T^{a_{\sigma(i)}}T^{a_j}T^{a_n}T^{a_{n+1}}T^{a_{\sigma(i+1)}}\ldots) = \Tr(\ldots T^{a_{\sigma(i)}}T^{a_j}T^{a_{n+1}}T^{a_n}T^{a_{\sigma(i+1)}}\ldots)\nonumber\\
	& \Tr(\ldots T^{a_{\sigma(i)}}T^{a_j}T^{a}T^{a}T^{a_{\sigma(i+1)}}\ldots)=(N)\Tr(\ldots T^{a_{\sigma(i)}}T^{a_j}T^{a_{\sigma(i+1)}}\ldots).
	\end{align}
	hese two ordering have the same group factor, which means their partial amplitudes can be combined together. To get the double poles, we first take the limit $z_{n+2}\to z_{n+1}$, then take the limit $z_{n+1}\to z_j$. It is easy to get the following poles from the diverging parts of the partial amplitudes
	\begin{equation}
	\label{eq:doublepole_1}
	\frac{1}{z_{j,n+1}z_{n+1,n+2}z_{n+2,\sigma(i+1)}}    + \frac{1}{z_{j,n+2}z_{n+2,n+1}z_{n+1,\sigma(i+1)}}  =\frac{1}{z_{j,n+1}^2 z_{j,\sigma(i+1)}} + \frac{2}{z_{j,n+1}z_{j,\sigma(i+1)}^2} ,
	\end{equation}
	where a single pole also arises in addition to the double pole. Combining the group factor and the remaining parts of the partial amplitudes, we get an MHV amplitude of $n$ gluons
	\begin{align}
	\label{eq:doubleorder_1}
	\lf(\frac{1}{z_{j,n+1}^2} + \frac{2}{z_{j,n+1}z_{j,\sigma(i+1)}}\ri)&N\Tr(\ldots T^{a_{\sigma(i)}}T^{a_j}T^{a_{\sigma(i+1)}}\ldots)\ \lf(\ldots\frac{1}{z_{\sigma(i),j}z_{j,\sigma(i+1)}}\ldots\ri)\nonumber\\
	&=\frac{N}{\omega_{n+1}\omega_{n+2}}\ \lf(\frac{1}{z_{j,n+1}^2} + \frac{2}{z_{j,n+1}z_{j,\sigma(i+1)}}\ri)\ A_{n}(\{\omega_i, q_i,J_i\}).
	\end{align}
	
%{\color{red} Define $M^{n}(\{\omega_i, q_i,J_i,a_i\})$, it should have a relation
%$A^{\sigma}_{J_1J_2\ldots J_{n}}\left(\{\omega_{i}, z_{i}, \bar{z}_{i}\}\right)$ in \req{eq:4d_amp}}
	
	For the other two orderings $(\sigma(i),n+1,n+2,j,\sigma(i+1))$ and $(\sigma(i),n+2,n+1,j,\sigma(i+1))$, it is easy to see that the contraction of color indices is the same as~\eqref{eq:color_1}. Then it is straightforward to get the following poles from the diverging parts of their partial amplitudes
	\begin{equation}
	\label{eq:doublepole_2}
	\frac{1}{z_{\sigma(i),n+1}z_{n+1,n+2}z_{n+2,j}}    + \frac{1}{z_{\sigma(i),n+2}z_{n+2,n+1}z_{n+1,j}}  =\frac{1}{z_{n+1,j}^2 z_{\sigma(i),j}} + \frac{2}{z_{n+1,j}z_{\sigma(i),j}^2}\ .
	\end{equation}
	So the result is also an MHV amplitude of $n$ gluons
	\begin{align}
	\label{eq:doubleorder_2}
	\lf(\frac{1}{z_{n+1,j}^2} + \frac{2}{z_{n+1,j}z_{\sigma(i),j}}\ri)&N\Tr(\ldots T^{a_{\sigma(i)}}T^{a_j}T^{a_{\sigma(i+1)}}\ldots)\ \lf(\ldots\frac{1}{z_{\sigma(i),j}z_{j,\sigma(i+1)}}\ldots\ri)\nonumber\\
	&=\frac{N}{\omega_{n+1}\omega_{n+2}}\lf(\frac{1}{z_{n+1,j}^2} + \frac{2}{z_{n+1,j}z_{\sigma(i),j}}\ri)\ A_{n}(\{\omega_i,q_i,J_i\}).
	\end{align}
	
	For the remaining two orderings  $(\sigma(i),n+1,j,n+2,\sigma(i+1))$ and $(\sigma(i),n+2,j,n+1,\sigma(i+1))$, the contraction of color indices has a single trace term $\Tr(T^{a_j})=0$, which is zero because it is a gluon. So the contribution is zero.
	
	Combining all the above results, we get the following result from all possible orderings in~\eqref{eq:pole_double_ordering}
	\begin{equation}
	\label{eq:doubleorder}
	\frac{2N}{\omega_{n+1}\omega_{n+2}}\ \lf(\frac{1}{z_{n+1,j}^2} + \frac{1}{z_{n+1,j}}\partial_j\ri)\ A_{n}(\{\omega_i, q_i,J_i\}),
	\end{equation}
	which contains a double pole and a single pole with a derivative acting on the $j$-th gluon.
	
	Now let us  see what would happen if the gauge group is chosen to be $SU(N)$, in which case the following formula is used
	\begin{equation}
	\label{eq:su_n}
	(T^a)^j_k (T^a)^s_l= \delta^j_l \delta^s_k-\frac{1}{N} \delta^j_k \delta^s_l\ .
	\end{equation}
	The extra term in this formula leads to an extra term $-(1/N)\Tr(\ldots T^{a_{\sigma(i)}}T^{a_j}T^{a_{\sigma(n+1)}}\ldots)$ in the color index contractions of all six kinds of ordering. When combining the partial amplitudes, this extra term cancels out, leaving the complete result the same as~\eqref{eq:doubleorder} . This is of course the $U(1)$ decoupling identity of standard YM, as expected.
	
\ \\
\noindent\underline{The single poles:}\\
	\label{sec:sub_single}

	\noindent	
	Now let us analyze the terms that only contribute  to single poles.  Consider the four orderings $(\sigma(i),j,n+1,\ldots,n+2,\sigma(i+1))$,  $(\sigma(i),n+1,j,\ldots,n+2,\sigma(i+1))$, $(\sigma(i),n+2,\ldots,j,n+1,\sigma(i+1))$ and $(\sigma(i),n+2,\ldots,n+1,j,\sigma(i+1))$. Using~\eqref{eq:group_formula}, the contraction of color indices contains double traces
	\begin{align}
	\label{eq:single_1}
	\Tr(\ldots T^{a_{\sigma(i)}}T^{a_j}T^{a}\ldots &T^{a}T^{a_{\sigma(i+1)}}\ldots) = \Tr(\ldots T^{a_{\sigma(i)}}T^{a}\ldots T^{a}T^{a_j}T^{a_{\sigma(i+1)}}\ldots)\nonumber\\
&=\Tr(\ldots)\Tr(\ldots T^{a_{\sigma(i)}}T^{a_j}T^{a_{\sigma(i+1)}}\ldots)\nonumber\\
\Tr(\ldots T^{a_{\sigma(i)}}T^{a}T^{a_j}\ldots &T^{a}T^{a_{\sigma(i+1)}}\ldots) = \Tr(\ldots T^{a_{\sigma(i)}}T^{a}\ldots T^{a_j}T^{a}T^{a_{\sigma(i+1)}}\ldots)\nonumber\\
&= \Tr(T^{a_j}\ldots)\Tr(\ldots T^{a_{\sigma(i)}}T^{a_{\sigma(i+1)}}\ldots)\ .
	\end{align}
	Combining the corresponding partial amplitudes, they cancel out in the coinciding limit $z_{n+2}=z_{n+1}$ and $z_{n+1}=z_{j}$, so the complete result is zero.
	
	Now let's see what would happen if the gauge group is chosen to be $SU(N)$, where the  contraction of color indices contains extra single traces
	\begin{align}
	\label{eq:single_su}
	\Tr(\ldots T^{a_{\sigma(i)}}T^{a_j}T^{a}\ldots &T^{a}T^{a_{\sigma(i+1)}}\ldots) =\Tr(\ldots T^{a_{\sigma(i)}}T^{a}T^{a_j}\ldots T^{a}T^{a_{\sigma(i+1)}}\ldots)\nonumber\\
&= (-\frac{1}{N})\Tr(\ldots T^{a_{\sigma(i)}}T^{a_j}\ldots T^{a_{\sigma(i+1)}}\ldots)\nonumber\\
\Tr(\ldots T^{a_{\sigma(i)}}T^{a}\ldots T^{a}&T^{a_j}T^{a_{\sigma(i+1)}}\ldots) = \Tr(\ldots T^{a_{\sigma(i)}}T^{a}\ldots T^{a_j}T^{a}T^{a_{\sigma(i+1)}}\ldots)\nonumber\\
&= (-\frac{1}{N})\Tr(\ldots T^{a_{\sigma(i)}}\ldots T^{a_{j}}T^{a_{\sigma(i+1)}}\ldots).
	\end{align}
	Again the combination of partial amplitudes for these extra terms cancels out  in the coinciding limit $z_{n+2}=z_{n+1}$ and $z_{n+1}=z_{j}$. So the result is the same as the case of $U(N)$, which is zero. Similarly, for the four kinds of ordering $(\sigma(i),j,n+2,\ldots,n+1,\sigma(i+1))$,  $(\sigma(i),n+2,j,\ldots,n+1,\sigma(i+1))$, $(\sigma(i),n+1,\ldots,j,n+2,\sigma(i+1))$ and $(\sigma(i),n+1,\ldots,n+2,j,\sigma(i+1))$, the complete result is zero. The remaining four kinds of ordering $(\sigma(i),n+2,n+1,\ldots,j,\sigma(i+1))$,  $(\sigma(i),n+1,n+2,\ldots,j,\sigma(i+1))$, $(\sigma(i),j,\ldots,n+2,n+1,\sigma(i+1))$ and $(\sigma(i),j,\ldots,n+1,n+2,\sigma(i+1))$ is also zero.

	Eventually, after combining all  permutations for the singular terms of the 4d MHV amplitude we get the result \req{OPEn} for $j=3,4,\ldots,n$ and
	 $\tilde C_2(G)=2N$ for $G=SU(N)$. We can generalize the above manipulations to negative helicity gluons $j=1, \, 2.$ Due to the cyclic structure of the denominator for the partial amplitude~\eqref{eq:partial_amp}, the poles in ~\eqref{eq:doubleorder_1} and ~\eqref{eq:doubleorder_2} are the same for $j=1,\, 2.$ The only difference is in the single pole term, where the numerator $z_{12}^4$ would modify the derivative term of ~\eqref{eq:doubleorder} as \req{tderi}.

\subsection{Gluon color sums and  operator products for general gauge group}\label{sec:colorsums}
\def\vev#1{\langle #1\rangle}
\def\f{\tilde f}

In this subsection we worked out color sums with insertion of a pair of gauge currents.
This generalizes the previous discussion for the case of general gauge group $G$. We will need to develop several  gauge group identities, some of which are novel and potentially useful for scattering amplitude computations in general.
We start with the color decomposition of an $n+2$--point gluon amplitude
\begin{align}
\label{startDDM}
A_{n+2}(\{p_i,J_j\})
&=\sum_{\sigma \in S_n}\f^{a_{n+2} b_{\sigma(1)} x_{1}} \f^{x_{1} b_{\sigma(2)} x_{2}}\ldots \f^{x_{n-1} b_{\sigma(n)} a_{n+1}}\nonumber\\
&\times A_{n+2}\left(n+2, \sigma\left(1,2,\ldots,n\right), n+1\right)\ ,
\end{align}
with the partial subamplitudes $A_{n+2}(\ldots)$.
The color decomposition is w.r.t. a $n!$--dimensional basis of subamplitudes $A_{n+2}\left(n+2, \sigma\left(1,2,\ldots,n\right), n+1\right)$ subject to the
DDM representation \cite{DelDuca:1999rs},  with the structure constants
$$\f^{abc}=-i\ \tr(T^a[T^b,T^c])\ ,$$
with $T^a$ generators in the fundamental representation.
Note, that \req{startDDM} is just an other representation of the color sum
\req{eq:4d_amp} for $n\ra n+2$.

We are interested in the pair of gluons $g_{n+1},g_{n+2}$ of positive helicity and  in \req{startDDM} we shall consider their  double soft limits\footnote{Note, that in the MHV case the double--soft limit \req{doublesoft} gives rise to an exact equation:
$$A_{n+2}\left((n+2)^+, \sigma\left(1,2,\ldots,n\right), (n+1)^+\right)=\fc{1}{\vev{n+1,n+2}}\  \fc{\vev{\si(n),\si(1)}}{\vev{n+2,\si(1)}\ \vev{\si(n),n+1}} A_{n}\left(\sigma\left(1,2,\ldots,n\right)\right)\ .$$
The latter just describes  splitting off all dependence on the gluons $g_{n+1},g_{n+2}$ from the remaining amplitude.  This is the method discussed in the previous subsection.} $p_{n+1},p_{n+2}\ra0$
\begin{align}
A_{n+2}\left((n+2)^+, \sigma\left(1,2,\ldots,n\right), (n+1)^+\right)&\ra \fc{\vev{n+1,\si(1)}}{\vev{n+1,n+2}\vev{n+2,\si(1)}}\ \nonumber\\
&\times \fc{\vev{\si(n),\si(1)}}{\vev{\si(n),n+1}\vev{n+1,\si(1)}}
\ A_{n}\left(\sigma\left(1,2,\ldots,n\right)\right)\ ,\label{doublesoft}
\end{align}
and perform the sum over the color indices $a_{n+1}$ and $a_{n+2}$, i.e.:
\be\label{Need}
\sum_{a_{n+1},a_{n+2}} \delta_{a_{n+1}a_{n+2}}\ \f^{a_{n+2} b_{\sigma(1)} x_{1}} \f^{x_{1} b_{\sigma(2)} x_{2}}\ldots \f^{x_{n-1} b_{\sigma(n)} a_{n+1}}\ .
\ee
Actually, the object \req{Need} is interesting on its own, since it appears in (planar) one--loop gluon amplitudes. It yields the group trace
$\Tr(T_A^{b_{\sigma(1)}} \ldots T_A^{b_{\sigma(n)}})$ with the gauge group generators
$T_A^a$ in the adjoint representation with $(T_A^a)_{bc}=-\tilde f^{abc}$.
E.g. for $n=3$ we have \cite{vanRitbergen:1998pn}:
\be\label{Groupv}
\sum_a \f^{ab_{1}x_1}  \f^{x_1b_{2}x_2} \f^{x_2b_{3}a}=-\h\ \tilde C_2(G)\ \f^{b_{1}b_{2}b_{3}}\ ,
\ee
with the invariant $\tilde C_2(G)\equiv \tilde C_2$ referring to the adjoint representation of the gauge group $G$, cf. also footnote \ref{grouptheory}. For general $n\geq 4$ we decompose \req{Need} into combinations of symmetric tensors and fewer numbers of structure constants.
This way for $n=4$ we have \cite{vanRitbergen:1998pn}:
\be
\label{Groupvi}
\f^{ab_{1}x_1}\f^{x_1b_{2}x_2}  \f^{x_2b_{3}x_3}\f^{x_3b_{4}a}=
\tilde d_A^{b_{1}b_{2}b_{3}b_{4}}+\fc{1}{6}\ \tilde C_2(G)\ \lf\{\f^{b_{1}b_{4}a}\f^{ab_{2}b_{3}}-\f^{b_{1}b_{2}a}\f^{ab_{3}b_{4}}\ri\}\ ,
\ee
with the symmetric invariant tensor $d_A$ given as
trace over symmetrized products of gauge group generators $T^a_A$:
\be
\tilde d_A:=\tilde d_A^{b_1b_2b_3b_4}=\fc{1}{4!}\sum_{\pi\in S_4}\Tr\lf(T_A^{a_{\pi(1)}}T_A^{a_{\pi(2)}}T_A^{a_{\pi(3)}}T_A^{a_{\pi(4)}}\ri)\ .
\ee
Furthermore, for $n=5$ we derive:
\begin{align}
\label{Groupvii}
&\f^{ab_{1}x_1} \f^{x_1b_{2}x_2}  \f^{x_2b_{3}x_3}\f^{x_3b_{4}x_4}\f^{x_4b_{5}a}=\nonumber\\
&-\fc{1}{12}\tilde{C}_2(G) \lf\{\f^{b_1b_2a}\f^{ab_3c}\f^{cb_4b_5}+\f^{b_2b_3a}\f^{ab_4c}\f^{cb_5b_1}+
\f^{b_5b_1a}\f^{ab_2c}\f^{cb_3b_4}-\f^{b_1b_4a}\f^{ab_3c}\f^{cb_2b_5}\ri\}\nonumber\\[2mm]
&-\h\ \lf\{ \f^{b_1b_5a}\tilde d_A^{ab_2b_3b_4}+\f^{b_3b_2a}\tilde d_A^{ab_1b_4b_5}+\f^{b_4b_2a}\tilde d_A^{ab_1b_3b_5}+\f^{b_4b_3a}\tilde d_A^{ab_1b_2b_5}\ri\}\ .
\end{align}
For general $n$ we have the relation of the following structure
\begin{align}
\f^{a b_{\sigma(1)} x_{1}} \f^{x_{1} b_{\sigma(2)} x_{2}}\ldots \f^{x_{n-1} b_{\sigma(n)} a}
&=\tilde d_A^{b_{\sigma(1)}b_{\sigma(2)}\ldots b_{\si(n)}}+\ldots+\nonumber\\
&+\tilde C_2(G)\ \lf\{\tilde f^{n-2} \ldots\ri\}
\end{align}

Let us now introduce celestial coordinates. With $\vev{ij}=(\omega_i\omega_j)^{1/2}z_{ij}$ the split factors in \req{doublesoft} can be expressed in terms of celestial coordinates as:
\begin{align}
A_{n+2}\Big((n+2)^+, \sigma\left(1,2,\ldots,n\right),& (n+1)^+\Big)\ra \fc{1}{\omega_{n+1}\omega_{n+2}}\ \lf(\fc{1}{z_{n+1}-z_{n+2}}+\fc{1}{z_{n+2}-z_{\si(1)}}\ri)\nonumber\\
&\times\lf(\fc{1}{z_{\si(n)}-z_{n+1}}+\fc{1}{z_{n+1}-z_{\si(1)}}\ri)
 A_{n}\left(\sigma\left(1,2,\ldots,n\right)\right)\ .\label{Doublesoft}
\end{align}
With these preparations we may compute the color sum \req{startDDM} supplemented by \req{Need} and \req{Doublesoft}. One important observation is that all the terms
$\tfrac{1}{z_{n+1}-z_{n+2}}$ cancel in the color sum. Therefore, we may safely take the limit $z_{n+2}\ra z_{n+1}$.
In the sequel the following universal functions will specify the color sum.
There is the function
\be\label{universal}
Z_0=\sum_{i=1}^n\fc{1}{(z_{n+1}- z_i)^2}\ ,
\ee
universal to all color orderings and an other function
\begin{align}
Z_{i_1,i_2,\ldots,i_n}&=\fc{1}{(z_{n+1}-z_{i_1})(z_{n+1}- z_{i_2})}+
\fc{1}{(z_{n+1}-z_{i_2})(z_{n+1}- z_{i_3})}\nonumber\\
&+\ldots+\fc{1}{(z_{n+1}-z_{i_n})(z_{n+1}- z_{i_1})}\ ,\label{rationalZ}
\end{align}
which sums over all neighbours of a given color ordering $(i_1,i_2,\ldots,i_n)$.

%%%%%%%%%%%%%%%%%%%%%%%%%%%%%%%%%%%%%%%%%%%%%%%%%%%%%%%%%%%
\ \\
\noindent\underline{$n=3$:}\\
For $n=3$ with \req{Groupv} in the limit $z_5\ra z_4$ we find:
\begin{align}
A_{5}(\{g_{5}^+,g_1,g_2,g_3,g_{4}^+\}) &=-\tilde C_2(G)\ \f^{b_1b_2b_3}\ A_3(1,2,3)\nonumber\\
&\times\fc{1}{\omega_{4}\omega_{5}}\fc{1}{ z_{14} z_{24} z_{34}}\
\lf\{     \fc{\ z_{12}\ z_{13}}{z_{14}}
+\fc{ z_{13}z_{23}}{ z_{34}} -\fc{ z_{12}\  z_{23}}{ z_{24}} \ri\}\ .
\end{align}
Then, e.g. for $ z_4\ra  z_1$ we have the expansion series:
\begin{align}\label{OPEi}
\lim_{ z_4\ra z_1}A_{5}(\{g_{5}^+,g_1,g_2,g_3,g_{4}^+\})
&=-\tilde C_2(G)\ \f^{b_1b_2b_3}\ A_3(1,2,3)\cr
&\times\fc{1}{\omega_{4}\omega_{5}}\lf\{\fc{1}{ z_{14}^2}+\fc{1}{ z_{14}}\lf(\fc{1}{ z_{12}}+\fc{1}{ z_{13}}\ri)\ri\}\ .
\end{align}
In the following we shall rewrite the subleading piece $\tfrac{1}{z_{14}}\lf(\ldots\ri)$ of \req{OPEi}. The three--point amplitudes are MHV amplitudes \req{eq:partial_amp}. Therefore, the latter assume the generic form
$$A_3(1,i_2,i_3)\sim \fc{z_{j_1j_2}^4}{z_{1i_2}z_{i_2i_3}z_{i_31}}\ ,$$
with $j_1,j_2$ denoting those two gluons of negative helicity.
After inspecting the rational terms in \req{OPEi} we deduce that the terms in the bracket
can be represented as derivative w.r.t. $z_1$ on the corresponding amplitude:
\begin{align}
\fc{1}{ z_{14}}\lf(\fc{1}{ z_{12}}+\fc{1}{ z_{13}}\ri)\ A_3(1,i_2,i_3)&=
 -\fc{1}{ z_{14}}\ \lf(\fc{\p}{\p z_1} -4\ \fc{\delta^{1,j_1}-\delta^{1,j_2}}{z_{j_1j_2}}\ri)\ A_3(1,i_2,i_3)\nonumber\\
&\equiv \fc{1}{ z_{41}}\ \tilde\p_{z_1}A_3(1,i_2,i_3)\ ,\label{receipt}
\end{align}
with the derivative \req{tderi}
singling out the two gluons $j_1,j_2$ with negative helicity.
Eventually, the limit \req{OPEi} gives rise to the following Ward identity:
\be
\lim_{ z_4\ra z_1}A_{5}(\{g_{5}^+,g_1,g_2,g_3,g_{4}^+\}) =-
\fc{\tilde C_2(G)}{\omega_{4}\omega_{5}}\lf(\fc{1}{ z_{41}^2}+\fc{\tilde\p_{z_1}}{ z_{41}}\ri)\  A_3(\{g_1,g_2,g_3\})\ ,
\ee
with the full three gluon amplitude:
\be\label{ThreeAmp}
A_3(\{g_1,g_2,g_3\})=\f^{b_1b_2b_3}\ A_3(1,2,3)\ .
\ee
Similar Ward identities can be derived for the other two cases $z_4\ra z_2,z_3$.
To this end, we get \req{OPEn} for $n=3$ with the amplitude  \req{ThreeAmp}.

%%%%%%%%%%%%%%%%%%%%%%%%%%%%%%%%%%%%%%%%%%%%%%%%%%%%%%%%%%%

\ \\
\noindent\underline{$n=4$:}\\
Next, for $n=4$ with \req{Groupvi} in the limit $z_6\ra z_5$ we determine:
\begin{align}
A_{6}(\{g_{6}^+,g_1,g_2,g_3,g_4,g_{5}^+\})&=\fc{1}{\omega_{5}\omega_{6}}\lf\{ \lf[\ \fc{\tilde C_2(G)}{3}\lf(-c_s-c_u\ri)+2d_A\ri] (Z_0-Z_{1234})\  A_4(1,2,3,4)\ri.\nonumber\\
&+\lf[\ \fc{\tilde C_2(G)}{3}\lf(-c_t+c_u\ri)+2d_A\ri] (Z_0-Z_{1324})\  A_4(1,3,2,4)\nonumber\\
&+\lf.\lf[\ \fc{\tilde C_2(G)}{3}\lf(c_s+c_t\ri)+2d_A\ri] (Z_0-Z_{1243})\  A_4(1,2,4,3)\ri\},\label{lastvi}
\end{align}
with the color factors
\begin{eqnarray}\label{FourPointColor}
c_s=\f^{b_1b_2a}\f^{b_3b_4a}\,, \hskip 0.8cm
c_t=\f^{b_1b_3a}\f^{b_2b_4a}\,, \hskip 0.8cm
c_u=\f^{b_4b_1a}\f^{b_2b_3a}\ ,
\end{eqnarray}
obeying  the Jacobi relation $c_t+c_u=c_s$.
Note, that the following Kleiss--Kuijf (KK) relation holds \cite{Kleiss:1988ne}:
\be\label{KK}
A_4(1,2,3,4)+A_4(1,2,4,3)+ A_4(1,3,2,4)=0\ .
\ee
As a consequence any universal term cancels in the above color sum \req{lastvi}.

Let us consider the limit $ z_5\ra  z_1$, for which we have:
\be\label{limits}
Z_0\ra \fc{1}{(z_5- z_1)^2}\ \ \ ,\ \ \ Z_\Sigma\ra \lf\{\begin{matrix}
\fc{1}{(z_1- z_5)} \lf(\fc{1}{ z_1- z_2}+\fc{1}{ z_1-z_4}\ri),&\Sigma=1234,\\[2mm]
\fc{1}{(z_1-z_5)} \lf(\fc{1}{ z_1- z_3}+\fc{1}{ z_1-z_4}\ri),&\Sigma=1324,\\[2mm]
\fc{1}{(z_1-z_5)} \lf(\fc{1}{ z_1- z_2}+\fc{1}{ z_1-z_3}\ri),&\Sigma=1243\ .
\end{matrix}    \ri.
\ee
Again in the  same way \req{receipt} as in the previous $n=3$ case we are able to rewrite the subleading pieces $\tfrac{1}{z_{15}}\lf(\ldots\ri)$  given in \req{limits} and entering \req{lastvi}. To this end as a consequence of \req{KK} up to the next leading order only the terms multiplying the color factors \req{FourPointColor} contribute in the color sum \req{lastvi}. The same conclusions can be drawn for the other limits  $z_5\ra z_2,z_3,z_4$ resulting in the following Ward identity \req{OPEn} with $n=4$ and the full four gluon amplitude:
\begin{align}
A_4(\{g_1,g_2,g_3,g_4\})&=\sum_{\sigma\in S_2}
\f^{b_1b_{\sigma(2)}a}\f^{ab_{\sigma(3)}b_4}\ A_4(1,\si(2),\si(3),4)\nonumber\\
&=c_s\ A_4(1,2,3,4)+c_t\ A_4(1,3,2,4)\ .
\end{align}

%%%%%%%%%%%%%%%%%%%%%%%%%%%%%%%%%%%%%%%%%%%%%%%%%%%%%%%%%%%

\noindent\underline{$n=5$:}\\
Next, for $n=5$ with \req{Groupvii} in the limit $z_7\ra z_6$ we derive:
$$\begin{array}{lcl}
&&A_{7}(\{g_{7}^+,g_1,g_2,g_3,g_4,g_5,g_{6}^+\})=-\fc{1}{\omega_{6}\omega_{7}}\nonumber \\[2mm]
&\times&\lf\{\lf[\fc{\tilde C_2(G)}{6}\lf(-c_1-c_2-c_5+c_6\ri)+x_1+x_2+x_3+x_5+x_6+x_8\ri] (Z_0-Z_{12345})\  A_5(1,2,3,4,5)\ri.\\[3mm]
&+&\lf[\fc{\tilde C_2(G)}{6}\lf(c_1-c_7+c_{11}+c_{14}\ri)-x_1+x_2+x_3+x_5+x_6+x_8\ri] (Z_0-Z_{12354})\  A_5(1,2,3,5,4)\\[3mm]
&+&\lf[\fc{\tilde C_2(G)}{6}\lf(c_5+c_9-c_{11}-c_{12}\ri)+x_1+x_2+x_3-x_5+x_6+x_8\ri] (Z_0-Z_{12435})\  A_5(1,2,4,3,5)\\[3mm]
&+&\lf[\fc{\tilde C_2(G)}{6}\lf(c_2+c_{12}+c_{15}-c_{10}\ri)+x_1-x_2+x_3-x_5+x_6+x_8\ri] (Z_0-Z_{12453})\  A_5(1,2,4,5,3)\\[3mm]
&+&\lf[\fc{\tilde C_2(G)}{6}\lf(c_2+c_{11}+c_{14}-c_{15}\ri)+x_1+x_2+x_3+x_5+x_6+x_8\ri] (Z_0-Z_{13245})\  A_5(1,3,2,4,5)\\[3mm]
&+&\lf[\fc{\tilde C_2(G)}{6}\lf(c_6+c_{7}+c_{15}-c_{5}\ri)-x_1+x_2+x_3+x_5+x_6-x_8\ri] (Z_0-Z_{13254})\  A_5(1,3,2,5,4)\\[3mm]
&+&\lf[\fc{\tilde C_2(G)}{6}\lf(c_5+c_{12}-c_{9}-c_{11}\ri)+x_1+x_2+x_3+x_5-x_6-x_8\ri] (Z_0-Z_{13425})\  A_5(1,3,4,2,5)\\[3mm]
&+&\lf[\fc{\tilde C_2(G)}{6}\lf(c_1+c_{3}+c_{9}-c_{2}\ri)+x_1+x_2-x_3+x_5-x_6-x_8\ri] (Z_0-Z_{13452})\  A_5(1,3,4,5,2)\\[3mm]
&+&\lf[\fc{\tilde C_2(G)}{6}\lf(c_2+c_{11}+c_{15}-c_{14}\ri)+x_1+x_2+x_3-x_5-x_6+x_8\ri] (Z_0-Z_{14235})\  A_5(1,4,2,3,5)\\[3mm]
&+&\lf[\fc{\tilde C_2(G)}{6}\lf(c_5+c_{9}+c_{10}+c_{14}\ri)+x_1-x_2+x_3-x_5-x_6+x_8\ri] (Z_0-Z_{14253})\  A_5(1,4,2,5,3)
\end{array}$$
\be\label{lastvii}
\begin{array}{lcl}
&+&\lf[\fc{\tilde C_2(G)}{6}\lf(c_1-c_{2}-c_{5}-c_{6}\ri)+x_1+x_2+x_3-x_5-x_6-x_8\ri] (Z_0-Z_{14325})\  A_5(1,4,3,2,5)\\[3mm]
&+&\lf.\lf[\fc{\tilde C_2(G)}{6}\lf(c_6+c_{12}-c_{3}-c_{11}\ri)+x_1+x_2-x_3-x_5-x_6-x_8\ri] (Z_0-Z_{14352}) A_5(1,4,3,5,2) \ri\},
\end{array}
\ee
with the  color factors \cite{Bern:2008qj}
\begin{eqnarray}
&&
c_{1\phantom{0}} = \f^{b_1 b_2 a}\f^{a b_3 c}\f^{c b_4 b_5}\,, \hskip 0.8cm
c_{2\phantom{1}} = \f^{b_2 b_3 a}\f^{a b_4 c}\f^{c b_5 b_1}\,, \hskip 0.8cm
c_{3\phantom{1}} = \f^{b_3 b_4 a}\f^{a b_5 c}\f^{c b_1 b_2}\,, \nonumber \\&&
c_{4\phantom{1}} = \f^{b_4 b_5 a}\f^{a b_1 c}\f^{c b_2 b_3}\,, \hskip 0.8cm
c_{5\phantom{1}} = \f^{b_5 b_1 a}\f^{a b_2 c}\f^{c b_3 b_4}\,, \hskip 0.8cm
c_{6\phantom{1}} = \f^{b_1 b_4 a}\f^{a b_3 c}\f^{c b_2 b_5}\,, \nonumber \\&&
c_{7\phantom{1}} = \f^{b_3 b_2 a}\f^{a b_5 c}\f^{c b_1 b_4}\,, \hskip 0.8cm
c_{8\phantom{1}} = \f^{b_2 b_5 a}\f^{a b_1 c}\f^{c b_4 b_3}\,, \hskip 0.8cm
c_{9\phantom{1}} = \f^{b_1 b_3 a}\f^{a b_4 c}\f^{c b_2 b_5}\,, \nonumber  \\&&
c_{10} = \f^{b_4 b_2 a}\f^{a b_5 c}\f^{c b_1 b_3}\,, \hskip 0.8cm
c_{11} = \f^{b_5 b_1 a}\f^{a b_3 c}\f^{c b_4 b_2}\,, \hskip 0.8cm
c_{12} = \f^{b_1 b_2 a}\f^{a b_4 c}\f^{c b_3 b_5}\,,  \nonumber \\ &&
c_{13} = \f^{b_3 b_5 a}\f^{a b_1 c}\f^{c b_2 b_4}\,, \hskip 0.8cm
c_{14} = \f^{b_1 b_4 a}\f^{a b_2 c}\f^{c b_3 b_5}\,, \hskip 0.8cm
c_{15} = \f^{b_1 b_3 a}\f^{a b_2 c}\f^{c b_4 b_5}\,. \hskip 1.5 cm
\label{FivePointColor}
\end{eqnarray}
fulfilling various Jacobi relations leaving the set of six independent $\{c_1,c_6,c_9,c_{12},c_{14},c_{15}\}$, and the ten tensors
\begin{eqnarray}
&&
x_{1\phantom{0}}=  \f^{b_4 b_5 c}d_A^{c b_1 b_2b_3}\,, \hskip 1cm
x_{2\phantom{0}}=  \f^{b_3 b_5 c}d_A^{c b_1 b_2b_4}\,, \hskip 1cm
x_{3\phantom{0}}=  \f^{b_2 b_5 c}d_A^{c b_1 b_3b_4}\,, \nonumber \\&&
x_{4\phantom{0}}=  \f^{b_1 b_5 c}d_A^{c b_2 b_3b_4}\,, \hskip 1cm
x_{5\phantom{0}}=  \f^{b_3 b_4 c}d_A^{c b_1 b_2b_5}\,, \hskip 1cm
x_{6\phantom{0}}=  \f^{b_2 b_4 c}d_A^{c b_1 b_3b_5}\,, \nonumber \\&&
x_{7\phantom{0}}=  \f^{b_1 b_4 c}d_A^{c b_2 b_3b_5}\,, \hskip 1cm
x_{8\phantom{0}}=  \f^{b_2 b_3 c}d_A^{c b_1 b_4b_5}\,, \hskip 1cm
x_{9\phantom{0}}=  \f^{b_1 b_3 c}d_A^{c b_2 b_4b_5}\,, \nonumber  \\&&
x_{10\phantom{0}}=  \f^{b_1 b_2 c}d_A^{c b_3 b_4b_5}\ ,
\label{FivePointTensor}
\end{eqnarray}
which fulfill the relations
\begin{eqnarray}
&&
x_{4\phantom{0}}= -x_1-x_2-x_3\,, \hskip 2cm
x_{7\phantom{0}}= x_1-x_5-x_6\,, \nonumber \\&&
x_{9\phantom{0}}= x_2+x_5-x_8\,, \hskip 2cm
x_{10\phantom{0}}= x_3+x_6+x_8\ ,
\label{TensorRel}
\end{eqnarray}
leaving six independent combinations $\{x_1,x_2,x_3,x_5,x_6,x_8\}$.

Again, for the limits $ z_6\ra  z_j,\ j=1,\ldots,5$ the leading term $\tfrac{1}{z_{j6}^2}$ can easily be extracted from \req{lastvii} by taking into account Jacobi and KK relations. To determine the next leading piece  $\tfrac{1}{z_{j6}}$ arising from \req{rationalZ} we can proceed in the  same way  as in the previous case $n=4$ which has lead to the receipt \req{receipt}. To this end we find \req{OPEn} for $n=5$ with the full five gluon amplitude:
\begin{align}
A_5(\{g_1,g_2,g_3,g_4,g_5\})&=\sum_{\sigma\in S_3}
\f^{b_1b_{\sigma(2)}a}\f^{ab_{\sigma(3)}c}\f^{cb_{\sigma(4)}b_5}\ A_5(1,\si(2),\si(3),\si(4),5)\nonumber\\
&=c_1\ A_5(1,2,3,4,5)+c_{12}\ A_5(1,2,4,3,5)+c_{15}\ A_5(1,3,2,4,5)\nonumber\\
&+c_9\ A_5(1,3,4,2,5)+c_{14}\ A_5(1,4,2,3,5)+c_{6}\ A_5(1,4,3,2,5)\ .
\end{align}

Finally, for generic $n$ we compute the color sum \req{startDDM} supplemented by \req{Need} and \req{Doublesoft} and consider the  limit $z_{n+2}\ra z_{n+1}$.
From the consideration above it is evident, that for general $n$ we obtain \req{OPEn}
with the full $n$ gluon amplitude:
\begin{align}
A_{n}(\{g_1,\ldots,g_{n}\})
&=\sum_{\sigma \in S_{n-2}}\f^{b_1b_{\sigma(2)} x_{1}} \f^{x_{1} b_{\sigma(3)} x_{2}}\ldots \f^{x_{n-5} b_{\sigma(n-3)} b_{n-2}}\nonumber\\
&\times A_{n}\left(1, \sigma\left(1,2,\ldots,n-1\right), n\right)\ .
\end{align}
This completes the general proof of equation (\ref{OPEn}). In the following section we will perform the Mellin transform of this amplitude and derive the OPE of the Sugawara energy-momentum tensor with primaries.

\section{Mellin transform and the Sugawara energy-momentum tensor}
	\label{sec:coincide_result}
In this section we will use the results we derived for gauge theory  amplitudes to derive the OPE of the Sugawara energy-momentum tensor with the operators of our theory. We will split the discussion into two parts. One part regarding the conformally soft gluons and in the second part we will discuss the hard states. We will see that although the Sugawara energy-momentum tensor has the right OPE to generate conformal transformations for  soft operators, it is not so for the hard ones and a modified tensor will be necessary. The correction needed will not be discussed in this work, although in section \ref{sec:shadow} we will propose an alternative construction, and it is an interesting open question.

\subsection{Sugawara energy-momentum tensor and conformally soft gluons}\label{sec:Sugasoft}	
We start with the OPE result \req{OPEn}  for generic gauge groups $G$.	After Mellin transforming the latter  we extract the OPE of the energy--momentum tensor $T^S(z_{n+1})$ with the currents $j^{a_j}(z_j)$ from the celestial amplitude~\eqref{eq:celestial_amp}. We follow the steps explained in the beginning of section~\ref{sec:Sugawara_ope}.
	The Mellin transform leads to
	\begin{align}
	\lim_{z_{n+1}\to z_j}\langle&\mathcal{O}_{\Delta_{1} J_{1}}^{a_{1}}(z_1,\bar{z}_1)\ldots \mathcal{O}_{\Delta_{j}+}^{a_{j}}(z_{j},\bz_j)\ldots\mathcal{O}_{\Delta_{n+1} +}^{a}(z_{n+1},\bz_{n+1}) \mathcal{O}_{\Delta_{n+2}+}^{a}(z_{n+1}, \bz_{n+1})\rangle\nonumber \\
	&=\lim_{z_{n+1}\to z_j}\left(\prod_{i=1}^{n+2} g\left(\Delta_{i}\right) \int\limits_0^\infty d \omega_{i}\ \omega_{i}^{i \lambda_{i}}\right)\ \times\delta^{4}\left(\sum_{i=1}^{n}\epsilon_{i}\omega_{i} q_{i}+\epsilon_{n+1}\omega'_{n+1}q_{n+1}\right) \nonumber\\
	&\times \frac{\tilde C_2(G)}{\omega_{n+1}\omega_{n+2}}\lf(\frac{1}{z_{n+1,j}^2} + \frac{\partial_j}{z_{n+1,j}}\ri)\ A_n(\{-,-,+,\ldots,+\})\ ,
	\label{eq:mellin_0}
	\end{align}
	where in the coinciding limit\footnote{Without losing generality, we assume that  collinear particles are either incoming or outgoing, i.e. $\epsilon_{j}=\epsilon_{n+1}=\epsilon_{n+2}$. In fact, we can assume that all positive helicity particles are outgoing as in reference~\cite{Fotopoulos:2019tpe}. } we can define the total energy $\omega'_{n+1}=\omega_{n+1}+\omega_{n+2}$ of the collinear pair.

Consider first the Mellin integral of the double pole part
	\begin{align}
	\label{eq:mellin_d0}
	\frac{\tilde C_2(G)}{z_{n+1,j}^2}&\lim_{z_{n+1}\to z_j}\left(\prod_{i=1}^{n+2} g\left(\Delta_{i}\right) \int_0^\infty d \omega_{i}\ \omega_{i}^{i \lambda_{i}}\right)
	\frac{1}{\omega_{n+1}\omega_{n+2}}\nonumber\\
	&\times\delta^{4}\left({\sum_{i=1\atop i\neq j}^n}\epsilon_{i}\omega_{i} q_{i}+\epsilon_j\omega'_jq_j\right)A_n(\{-,-,+,\ldots,+\})\ ,
	\end{align}
	where in the coinciding limit $z_{n+1}=z_j$ we can further define $\omega'_{j}=\omega_j+\omega'_{n+1}$.
	The integral of the collinear states becomes
	\begin{align}
	\label{eq:int_d0}
	&\int_0^\infty d\omega_j d\omega_{n+1} d\omega_{n+2}\,  \omega_j^{i\lambda_j}\omega_{n+1}^{i\lambda_{n+1}}\omega_{n+2}^{i\lambda_{n+2}} \frac{1}{\omega_j \omega_{n+1}\omega_{n+2}}\ldots \nonumber\\
	&=\int_0^\infty d\omega'_j\int_0^{\omega'_j} d\omega'_{n+1} \int_0^{\omega'_{n+1}} d\omega_{n+1}\, \omega_{n+1}^{-1+i\lambda_{n+1}} (\omega'_{n+1} - \omega_{n+1})^{-1+i\lambda_{n+2}}(\omega'_j - \omega'_{n+1})^{-1+i\lambda_{j}}\ldots  \nonumber\\
	&=B(i\lambda_{n+1},i\lambda_{n+2})B(i\lambda'_{n+1},i\lambda_{j})\int_0^\infty d\omega'_j {\omega'}_j^{-1+i\lambda'_j} \ldots,
	\end{align}
	where we use the new variables $\lambda'_{n+1}=\lambda_{n+1}+\lambda_{n+2}$ and $\lambda'_j=\lambda_j+\lambda'_{n+1}$.
	Combing with the normalization factors $g(\lambda_j), g(\lambda_{n+1}), g(\lambda_{n+2})$ and taking the conformal soft limit $\Delta_{n+1}=\Delta_{n+2}=1$, we obtain the double pole of the OPE
\begin{equation}
	\label{eq:ope_d0}
	\frac{\tilde C_2(G)}{z_{n+1,j}^2} \  \langle\mathcal{O}_{\Delta_{1} J_{1}}^{a_{1}}\ldots \mathcal{O}_{\Delta_{j}+}^{a_{j}}(z_{j})\ldots\rangle\ .
	\end{equation}

Next, for the Mellin integral of the single pole part, we can move the derivative out of the integral by adding an extra term:
	\begin{align}
	\label{eq:mellin_single}
	&\frac{\tilde C_2(G)}{z_{n+1,j}}\lim_{z_{n+1}\to z_j}\partial_j\left\{\left(\prod_{i=1}^{n+2} g\left(\lambda_{i}\right) \int d \omega_{i} \omega_{i}^{i \lambda_{i}}\right)\delta^{4}\left(\sum_{i=1}^{n}\epsilon_{i}\omega_{i} q_{i}+\epsilon_n\omega'_{n+1}q_{n+1}\right)\frac{ A_n(\{\ldots\})}{\omega_{n+1}\omega_{n+2}}\right\} \nonumber\\
	&-\frac{\tilde C_2(G)}{z_{n+1,j}}\lim_{z_{n+1}\to z_j}\left(\prod_{i=1}^{n+2} g\left(\lambda_{i}\right) \int d \omega_{i} \omega_{i}^{i \lambda_{i}}\right)\frac{A_n(\{\ldots\})}{\omega_{n+1}\omega_{n+2}}  \left\{\partial_j\delta^{4}\left(\sum_{i=1}^{n}\epsilon_{i}\omega_{i} q_{i}+\epsilon_{n+1}\omega'_{n+1}q_{n+1}\right)\right\}.
	\end{align}
	For the second term in~\eqref{eq:mellin_single}, the coinciding limit $z_{n+1}=z_j$ and the derivative $\partial_j$ on the delta function do not commute. An explicit computation shows that
	\begin{equation}
	\label{eq:delta_derivative}
	\lim_{z_{n+1}\to z_j}\partial_j\delta^{4}\left(\sum_{i=1}^{n}\epsilon_{i}\omega_{i} q_{i}+\epsilon_{n+1}\omega'_{n+1}q_{n+1}\right) =
\frac{\omega_j}{\omega'_j}\partial_j\delta^{4}\left(\sum_{i=1\atop i\neq j}^n\epsilon_{i}\omega_{i} q_{i}+\epsilon_j\omega'_jq_j\right),
	\end{equation}
	where on the right hand side the coinciding limit $z_{n+1}=z_j$ was used and we defined  $\omega'_j=\omega_j+\omega'_{n+1}$.
	For the first term in~\eqref{eq:mellin_single}, using the delta function  expanded around $z_{n+1}=z_j+z_{n+1,j}$
	\begin{align}
	\label{eq:delta_expand}
	&\delta^{4}(\sum_{i=1}^{n}\epsilon_{i}\omega_{i} q_{i}+\epsilon_{n+1}\omega'_{n+1}q_{n+1}) \nonumber\\
	&= \delta^{4}(\sum_{i=1\atop i\neq j}^n\epsilon_{i}\omega_{i} q_{i}+\epsilon_j\omega'_jq_j) + z_{n+1,j}
	\partial_{n+1}\delta^{4}(\sum_{i=1}^{n}\epsilon_{i}\omega_{i} q_{i}+\epsilon_n\omega'_{n+1}q_{n+1})|_{z_{n+1}=z_j}\nonumber\\
	&= \delta^{4}(\sum_{i=1\atop i\neq j}^n\epsilon_{i}\omega_{i} q_{i}+\epsilon_j\omega'_jq_j) + z_{n+1,j}\frac{\omega'_{n+1}}{\omega'_j}\partial_j\delta^{4}(\sum_{i=1\atop i\neq j}^n\epsilon_{i}\omega_{i} q_{i}+\epsilon_j\omega'_jq_j)
	\end{align}
we get:
	\begin{align}
	\label{eq:mellin_single_1st}
	\frac{\tilde C_2(G)}{z_{n+1,j}}\lim_{z_{n+1}\to z_j}\partial_j\int&\left\{  \delta^{4}(\sum_{i=1\atop i\neq j}^n\epsilon_{i}\omega_{i} q_{i}+\epsilon_j\omega'_jq_j) + z_{n+1,j}\frac{\omega'_{n+1}}{\omega'_j}\partial_j\delta^{4}(\sum_{i=1\atop i\neq j}^n\epsilon_{i}\omega_{i} q_{i}+\epsilon_j\omega'_jq_j) (\ldots)\right\} \nonumber\\
	&=\frac{\tilde C_2(G)}{z_{n+1,j}}\partial_j\int\delta^{4}(\sum_{i=1\atop i\neq j}^n\epsilon_{i}\omega_{i} q_{i}+\epsilon_j\omega'_jq_j)  (\ldots)\nonumber\\
	& -\frac{\tilde C_2(G)}{z_{n+1,j}}\int\frac{\omega'_{n+1}}{\omega'_j}\partial_j\delta^{4}(\sum_{i=1\atop i\neq j}^n\epsilon_{i}\omega_{i} q_{i}+\epsilon_j\omega'_jq_j) (\ldots) +\mathcal{O}(z_{n+1,j}^0).
\end{align}
	Combining the two terms eq.~\eqref{eq:delta_derivative} and eq.~\eqref{eq:mellin_single_1st}, the Mellin transform~\eqref{eq:mellin_single} becomes
	\begin{align}
	\label{eq:int_single}
	&\frac{\tilde C_2(G)}{z_{n+1,j}}\partial_j\left\{\left(\prod_{i=1}^{n+2} g\left(\lambda_{i}\right) \int d \omega_{i} \omega_{i}^{i \lambda_{i}}\right)\frac{1}{\omega_{n+1}\omega_{n+2}}\delta^{4}(\sum_{i=1\atop i\neq j}^n\epsilon_{i}\omega_{i} q_{i}+\epsilon_j\omega'_jq_j)\ A_n(\{\ldots\})\ \right\} \nonumber\\
	&-\frac{\tilde C_2(G)}{z_{n+1,j}}\left(\prod_{i=1}^{n+2} g\left(\lambda_{i}\right) \int d \omega_{i} \omega_{i}^{i \lambda_{i}}\right)\frac{1}{\omega_{n+1}\omega_{n+2}}\ A_n(\{\ldots\})\ \left\{\partial_j\delta^{4}(\sum_{i=1\atop i\neq j}^n\epsilon_{i}\omega_{i} q_{i}+\epsilon_j\omega'_jq_j)\right\}.
	\end{align}
	The energy integral over $\omega_{n+1},\omega_{n+2}$ is the same as the integral of the double pole part~\eqref{eq:int_d0}, so this part of the OPE is
	\begin{align}
	\label{eq:ope_s0}
	\frac{\tilde C_2(G)}{z_{n+1,j}} \bigg(\partial_j \langle\mathcal{O}_{\Delta_{1} J_{1}}^{a_{1}}\ldots \mathcal{O}_{\Delta_{j}+}^{a_{j}}(z_{j})\ldots\rangle-\big(g(\lambda_j)\int_0^\infty d\omega_j \omega_j^{i\lambda_j}\ldots\big)\ A_n(\{\ldots\})\  \partial_j\delta^{4}(\sum_{i=1}^{n}\epsilon_{i}\omega_{i} q_{i})\bigg),
	\end{align}
	where the integration variable is changed from $\omega'_j$ to $\omega_j$ for convenience.
	
	As discussed in section \ref{sec:Prelim}, the Sugawara energy-momentum tensor is expected to describe the conformal properties of the soft sector of the theory, which corresponds to the conformal soft limit~\cite{fan_soft_2019}. Therefore as explained earlier, we need to take the conformal soft limit $\Delta_j=1$ of the $j$-th conformal field operator.  In the second term of~\eqref{eq:ope_s0}, the derivative $\partial_j$ on the delta function contributes an extra $\omega_j$ to the Mellin integral
	\begin{equation}
	\label{eq:delta_j}
	\partial_j\delta^{4}(\sum_{i=1}^{n}\epsilon_{i}\omega_{i} q_{i}) = \epsilon_j \omega_j (\partial_jq_j) \frac{\delta^{4}(\sum_{i=1}^{n}\epsilon_{i}\omega_{i} q_{i})}{\sum_{i=1}^{n}\epsilon_{i}\omega_{i} q_{i}}.
	\end{equation}
	According to the analysis~\cite{fan_soft_2019}, this extra $\omega_j$ gives a Mellin integral without a pole $1/\lambda_j$ so under the $\lambda_j \to 0$  conformal soft limit vanishes due to $g(\lambda_j)\to0$.
	
	Combining the above results, we obtain:
	\begin{align}
	\lim_{z_{n+1}\to z_j}&\langle\mathcal{O}^{a_1}_{\Delta_{1} ,J_1} \ldots j^{a_{j}}(z_{j})\ldots \mathcal{O}^{a_n}_{\Delta_{n} ,J_n}    j^{a}(z_{n+1}) j^{a}(z_{n+1})\rangle =
	\nonumber \\
&=\tilde C_2(G)\ 	\lf(\frac{1}{z_{n+1,j}^2} + \frac{\partial_j}{z_{n+1,j}} \ri) \langle\mathcal{O}^{a_1}_{\Delta_{1} ,J_1} \ldots j^{a_{j}}(z_{j})\ldots \mathcal{O}^{a_n}_{\Delta_{n} ,J_n}  \rangle\ , \qquad j=3,\ldots n\ .\label{eq:ope_f0}
	\end{align}
We see that the overall constant of the OPE is $\tilde C_2(G)$. Therefore, in (\ref{eq:sugawara_def}) we need to choose a normalization  $\gamma={1\over \tilde C_2(G)}$.
	%But, the fact is that the factor $2$ comes from the OPE of $t(z)j(w)$, and the normalization constant of the Sugawara $t(z)$ is $N$
We can define therefore,
	\begin{equation}
	\label{eq:sugawara_norm}
	T^S(z)=\frac{j^a(z)j^a(z)}{ \tilde C_2(G)}\ ,
	\end{equation}
	which  agrees with its general definition (\ref{eq:Sugawara_1})~\cite{DiFrancesco:1997nk} for level $k=0$.
	We have shown that indeed  the OPE of  $T^S(z)$ with a current $j^a(w)$ is given by:
	\begin{equation}
	\label{eq:sugawara_ope}
	T^S(z)j^a(w)=\frac{h}{(z-w)^2}j^a(w) + \frac{\partial_wj^a(w)}{z-w},\quad h=1 .
	\end{equation}
	%which agrees with the conformal soft limit of the result of the reference~\cite{Fotopoulos:2019tpe}.
	In  section \ref{sec:gengroup} we discuss the Sugawara construction and its OPE with the currents $j^a(z)$ for a general group, using directly the soft theorem for conformally soft states~\cite{fan_soft_2019}. It agrees with our result from the previous subsections, namely (\ref{eq:sugawara_ope}).
	
We close this subsection, with a comment for the case of the negative helicity gluons $j=1,2$ in eq.~\eqref{eq:ope_f0}. For  $j=1,2$, the respective Mellin integral in eq.~\eqref{eq:int_d0} and eq.~\eqref{eq:ope_d0}  will give the following result for the $j$-th operator
\begin{align}
\label{eq:tbarj}
g(\lambda_{j})\int_0^\infty d\omega_j\ {\omega}_j^{1+i\lambda_j} \ldots\overset{\lambda_j=0}{=} 0\ \ \ ,\ \ \ j=1,2\ .
\end{align}
It is zero under the conformal soft limit of the $j$-th operator, because there is no $1/\lambda_j$ pole~\cite{fan_soft_2019,Pate:2019mfs,Nandan:2019jas} that can cancel the $\lambda_j$ factor in $g(\lambda_j)$. Hence, for the case of MHV amplitudes $A_n(-,-,+,+,\ldots,+)$,  the conformal soft limit of the negative helicity states gives zero. This was observed also in \cite{Pate:2019mfs}, which is consistent with the vanishing of the MHV amplitude (\ref{eq:partial_amp}) in momentum space under the soft limit of a negative helicity gluon.
In total we can write \eqref{eq:ope_f0} as
\begin{align}
	\lim_{z_j\rightarrow z_{n+1}} &\langle\mathcal{O}_{\Delta_{1} J_{1}}^{a_{1}}\ldots j^{a_{j}}(z_{j})\ldots \mathcal{O}_{\Delta_{n} J_{n}}^{a_{n}}   T^S(z_{n+1})\rangle\nonumber\\
	&=\begin{cases}
	\displaystyle{\left( \frac{1}{(z_{n+1}\!-\!z_j)^2} \!+\! \frac{\partial_j}{(z_{n+1}-z_j)} \right) \langle\mathcal{O}_{\Delta_{1} J_{1}}^{a_{1}}\ldots j^{a_{j}}(z_{j})\ldots\mathcal{O}_{\Delta_{n} J_{n}}^{a_{n}} \rangle\ ,}\  &\ j=3,\ldots,n,\\
		\displaystyle{\ 0\ ,}\ &\ j=1,2\ ,	
	\end{cases}
\end{align}
with $\Delta_j\ra1$.

Therefore we cannot extract any OPE of $T^S(z)$ with $\bar{j}(\bw)$. Naively, we expect this OPE to be regular since the operators $\bar{j}(\bw)$ are antiholomorphic with weights $h=0$ and $\bh=1$. Also this assertion, although discussed here only for MHV amplitudes, makes a connection with our earlier discussion regarding the CS interpretation of the theory. Only one set of currents can survive in the soft limit, holomorphic or antiholomorphic. We have made the choice which leads to a holomorphic Kac-Moody algebra and the antiholomorphic currents $\bar{j}(\bw)$ are expected to decouple. Indeed, as shown above, this is the case for MHV amplitudes. But unfortunately this does not hold for $NMHV$ amplitudes, see appendix  \ref{sec:appendixB}. So we must restrict our discussion solely on correlators which involve only one type of soft gluons, positive or negative ones.  In appendix \ref{sec:appendixC} we discuss the role of the shadow transform, which allows soft negative helicity states to be expressed as positive helicity ones. Therefore allowing us to have a purely holomorphic correlator, alas in an apparently non-local formulation.

As a final remark, had we chosen to work in a $\overline{ MHV}$ basis we would be led to an antiholomorphic Kac-Moody algebra. For an antiholomorphic Kac-Moody algebra, we will get the $\bar{T}^S(\bar{z})\bar{j}^a(\bar{w})$-OPE as:
	\begin{equation}
	\label{eq:sugawara_Cope}
	\bar{T}^S(\bar{z})\bar{j}^a(\bar{w})=\frac{\bar{h}}{(\bar{z}-\bar{w})^2}\bar{j}^a(\bar{w}) + \frac{\partial_{\bar{w}}\bar{j}^a(\bar{w})}{\bar{z}-\bar{w}},\quad \bar{h}=1\ .
	\end{equation}
Similar conclusions as for the holomorphic sector apply in this case.

\subsection{Comments on the OPE of the Sugawara tensor for hard operators}\label{sec:comments}

On the other hand, hard operators $O^a_{\D,-}$ as well as $O^a_{\D, +}$ act as color sources for soft modes.  The complete theory requires an energy-momentum tensor for the hard states as well. We can try to examine the collinear limit of the Sugawara tensor with a hard operator. Looking carefully at (\ref{OPEn}) we see that the single poles for negative helicity states have a modified partial derivative (\ref{tderi}). This already poses an issue with the negative helicity gluons. Also, in the Mellin transform derivation we encounter (\ref{eq:ope_s0}). We see that only for the soft limit $\lambda_j \to 0$ we can recover the simple partial derivative of the celestial amplitude. Finally, the double poles for hard operators pose a problem as well. These states have weights $h=i{\lambda\over 2}$ and $\bar{h}= 1+i {\lambda\over 2}$. But the Sugawara energy momentum tensor will give always weights proportional to the eigenvalues of the quadratic Casimir operator on the states.  In all situations completing the Mellin integrals  and taking the double conformal soft limit for gluons $n+1,n+2$, results in the double poles of  (\ref{eq:sugawara_ope}). From the analysis above we conclude, that when the Sugawara energy momentum tensor acts on  hard negative  or hard positive  helicity states we do not derive the desired OPE.

One might wonder if we can consider a subsector of the theory where the Sugawara decouples completely from hard and soft negative helicity states. We know that all positive helicity states can be taken soft under consecutive soft limits. In the previous section we concluded though, that soft limits of negative helicity gluons, lead to vanishing MHV amplitudes. It is also known that a pure plus helicity amplitude does not exist in the gauge theory side and therefore in the CCFT we cannot have a correlator with only $j^a(z)$ operators \footnote{See though appendix \ref{sec:appendixC} for a formulation which includes shadow operators $\widetilde{j}^a(z)$.}.
 We conclude that we need to modify the Sugawara tensor to account for the proper conformal properties of hard operators.
As  suggested in \cite{Cheung:2016iub,McLoughlin:2016uwa} one should include an additional term in the definition of the full energy-momentum tensor:
\be\label{eq:Tful}
 T(z)= T^S(z) +T'(z)\ .
 \ee
 This remains an open problem. So at this stage, in order to have non-vanishing correlators with $j^a(z)$ operator insertions, we need correlators with heavy states of the CCFT.  As suggested in  \cite{He:2015zea,Cheung:2016iub,Nande:2017dba} we can treat external, heavy negative and positive helicity states as Wilson lines. In section \ref{sec:shadowKLT} we will make an alternative  proposal based on our analysis for the Einstein-Yang-Mills theory, which works for soft and hard operators alike.

 As an example we can consider massive particles as sources of soft gauge radiation  \cite{Nande:2017dba}. Massive particles are described by time-like Wilson lines which source soft gauge bosons. Unlike massless particles whose wave function localizes on the ${\cal C S}^2$ at null infinity, massive particles' trajectories do not asymptote to the celestial sphere ${\cal C S}^2$ at null infinity. Massless particles i.e gluons  correspond to local operators on  ${\cal C S}^2$, but massive particles correspond to smeared operators and involve non-local integrals of local operators on  ${\cal C S}^2$.

 To make our point we restrict to the case of QED to avoid a heavy notation with colour matrices and traces. For QED, consider as in \cite{Nande:2017dba}, the CCFT operator $O(p)$, where $p$ the four-momentum,  describes massive states. We will assume that this operator factorizes into two parts. One part  $\hat{O}(p)$ is neutral under large gauge transformations and decouples from soft radiation. The second part ${\cal W}_{Q}(p)$, where $Q$ is the charge of the massive state, is a Wilson line, a smeared operator on  ${\cal C S}^2$, that transforms under large gauge transformations and describes the coupling of massive states to soft radiation. Correlation functions on the CCFT are expected to factorize \cite{Feige:2014wja,Nande:2017dba}:
 \be\label{eq:factor}
 \langle j_1,j_2\ldots j_n O_1,O_2\ldots O_m\rangle= \underbrace{\langle \hat{O}_1,\hat{O}_2\ldots \hat{O}_m\rangle}_{hard}\underbrace{\langle j_1,j_2\dots j_n {\cal W}_{Q_1},{\cal W}_{Q_2}\ldots{\cal W}_{Q_m}\rangle}_{soft}\ .
 \ee
 The Sugawara tensor is expected to be the energy momentum tensor of the  sub-CFT of conformally soft operators and Wilson lines of the CCFT. It is very interesting to extend our present discussion in the non-Abelian case and include Wilson operators in our correlators, staring from the  hard-soft-collinear factorization of scattering amplitudes in QCD \cite{Feige:2014wja}.

\section{The general gauge group using OPE and conformal soft limits of currents}\label{sec:gengroup}

In this section we discuss the Sugawara energy--momentum tensor for general gauge groups by   using directly the conformal soft limit result of~\cite{fan_soft_2019} and subsequently the collinear limit. This is different sequence of operations compared to the one of the previous section. It can be applied only for the OPE of the energy momentum tensor with a soft gauge boson.  We still consider the MHV case and follow closely the discussion in chapter 15 of  \cite{DiFrancesco:1997nk}.
We write the Sugawara energy momentum tensor in the the form  (\ref{eq:Sugawara_1}). We will consider the following expression
\begin{align}
\label{eq:geng_1}
 {\gamma\over 2 \pi i} & \oint_{z_{n+2}} \ {d z_{n+1}\over z_{n+1}-z_{n+2}}\  \langle \mathcal{O}_{\Delta_{1} -}^{b_{1}}(z_1,\bar{z}_1) \ldots j^{b_n}(z_n) j^a(z_{n+1}) j^a(z_{n+2}) \rangle \nonumber \\
=&\lim_{\D_{n+1}\D_{n+2} \to 1} \ \lim_{\D_{n} \to 1} {\gamma\over 2 \pi i}  \oint_{z_{n+2}} \ {d z_{n+1}\over z_{n+1}-z_{n+2}}\  \langle \mathcal{O}_{\Delta_{1} -}^{b_{1}}(z_1,\bar{z}_1) \mathcal{O}_{\Delta_{2} -}^{b_{2}}(z_2,\bar{z}_2) \mathcal{O}_{\Delta_{3} +}^{b_{3}}(z_3,\bar{z}_3) \ldots \nonumber \\
&\mathcal{O}_{\Delta_{n} +}^{b_{n}}(z_n,\bar{z}_n)\mathcal{O}_{\Delta_{n+1} +}^{a}(z_{n+1},\bar{z}_{n+1})  \mathcal{O}_{\Delta_{n+2} +}^{a}(z_{n+2},\bar{z}_{n+2}) \rangle \ ,
\end{align}
where $\gamma$ a normalization constant to be determined soon.
Now we apply the soft limit iteratively following closely the derivation in equations (15.51-15.56) of \cite{DiFrancesco:1997nk}. We use the collinear and soft limits in (\ref{eq:JJope}). First, we take the soft limit for the operator $\mathcal{O}_{\Delta_{n} +}^{b_{n}}(z_{n},\bar{z}_{n})$
\begin{align}\label{eq:geng_2}
=&\sum_{i=1}^{n-1} \frac{\tilde{f}^{b_{n} b_i c}}{z_{n,i}} \left\langle \mathcal{O}_{\D_1}^{b_{1}}(z_1,\bar{z}_1) \ldots\mathcal{O}_{\D_i}^{c}(z_i,\bar{z}_i) \ldots \ri.\nonumber\\
&\lf.\ldots\mathcal{O}^{b_{n-1}}(z_{n-1},\bar{z}_{n-1})\mathcal{O}_{\Delta_{n+1} +}^{a}(z_{n+1},\bar{z}_{n+1}) \mathcal{O}_{\Delta_{n+2} +}^{a}(z_{n+1},\bar{z}_{n+1})\right\rangle \nonumber \\
&+ \frac{\tilde{f}^{b_{n} a c}}{z_{n,n+1}} \left\langle\mathcal{O}_{\Delta_{1}}^{b_{1}}(z_1,\bar{z}_1)\ldots\mathcal{O}_{\Delta_{n+1},+}^{c}(z_{n+1},\bar{z}_{n+1})\mathcal{O}_{\Delta_{n+2} +}^{a}(z_{n+2},\bar{z}_{n+2})\right\rangle \nonumber \\
 &+\frac{\tilde{f}^{ b_{n} a c}}{z_{n,n+2}} \left\langle\mathcal{O}_{\Delta_{1}}^{b_{1}}(z_1,\bar{z}_1)\ldots\mathcal{O}_{\Delta_{n+1},+}^{a}(z_{n+1},\bar{z}_{n+1})\mathcal{O}_{\Delta_{n+2} +}^{c}(z_{n+2},\bar{z}_{n+2})\right\rangle \ .
\end{align}
Then we consider the collinear limit for the operator $\mathcal{O}_{\Delta_{n} +}^{b_{n}}(z_n,\bar{z}_n)$ as it approaches the two operators  at $z_{n+1}$ and $z_{n+2}$
The first line above does not contribute to the OPE of interest since it is finite as $z_n\to z_{n+1}$. The last two lines are inserted in the contour integral (\ref{eq:geng_1}). Now we apply the conformal soft limit on the operator $\mathcal{O}^{a}_{\Delta_{n+1} ,+}\to J^a(z_{n+1})$ and use the OPE (\ref{eq:JJope}) with $\mathcal{O}^b_{\Delta_{n+2} ,+}$ .
\begin{align}\label{eq:geng_2b}
& {\gamma\over 2 \pi i}  \oint_{z_{n+2}} \ {d z_{n+1}\over z_{n+1,n+2}}\ \Big( \frac{\tilde{f}^{b_{n} a c}\tilde{f}^{c a d}}{z_{n,n+1}z_{n+1,n+2}} \left\langle\mathcal{O}_{\Delta_{1}}^{b_{1}}(z_1,\bar{z}_1)\ldots \mathcal{O}_{\Delta_{n+2} +}^{d}(z_{n+2},\bar{z}_{n+2})\right\rangle \nonumber \\
 & +\frac{\tilde{f}^{b_{n} a c}\tilde{f}^{ac  d}}{z_{n,n+2}z_{n+1,n+2}} \left\langle\mathcal{O}_{\Delta_{1}}^{b_{1}}(z_1,\bar{z}_1)\ldots \mathcal{O}_{\Delta_{n+2} +}^{d}(z_{n+2},\bar{z}_{n+2})\right\rangle \Big) .
\end{align}
The second term gives only regular terms in the contour integral. Only the first one is relevant. We use the integration formula:
\be\label{eq:integral}
{1\over 2 \pi i}  \oint_{w} {d x \over (x-w)^n} {F(w)\over (z-x)^m}= {(n+m-2)!\over (n-1)! (m-1)!} {F(w)\over (z-w)^{n+m-1}}\ .
\ee
Finally, we derive the following expression
\begin{align} \label{eq:geng_3}
	\gamma \  {-\tilde{C}_2\over (z_n-z_{n+2})^2} \left\langle\mathcal{O}_{\Delta_{1}}^{b_{1}}(z_1,\bar{z}_1)\ldots \mathcal{O}^{b_n}_{\Delta_{n+2} ,+}(z_{n+2},\bz_{n+2})\right\rangle,
\end{align}
where  the overall minus sign comes from the formula $\tilde{f}^{b_n ac}\tilde{f}^{c ad}=-\tilde{C}_2\delta^{b_nd}$ and will be thrown away because it reflects only the antisymmetric property of the structure  constant.
At this point we need to expand the correlator for $z_{n+2}$ around $z_n$. We can use a similar method as in \cite{Fotopoulos:2019tpe} or use the delta function expansion in (\ref{eq:delta_expand}) leading to (\ref{eq:ope_s0}).  Finally, lest consider the conformal soft limit of $\mathcal{O}_{\Delta_{n+2} ,+}$ and follow the discussion that leads to (\ref{eq:ope_f0}). In this situation, for the MHV case discussed in the previous section, we can simply Taylor expand the Mellin transform of the partial amplitude (\ref{eq:partial_amp})
At the end we arrive at
\begin{align} \label{eq:geng_4}
\left\langle\mathcal{O}_{\Delta_{1}}^{b_{1}}(z_1,\bar{z}_1)\ldots  T(z_{n+2}) j^{b_n}(z_{n})\right\rangle\sim  \gamma \ \tilde{C}_2\ \lf({1\over z_{n+2,n}^2} + \frac{\partial_n}{z_{n+2,n}} \ri) \left\langle\mathcal{O}_{\Delta_{1}}^{b_{1}}\ldots j^{b_n}(z_{n})\right\rangle\ ,
\end{align}
where the choice $\gamma={1\over \tilde{C}_2}$ gives the correct normalization for a level $k=0$ Sugawara energy momentum tensor. This concludes the derivation of the OPE for the Sugawara tensor using an approach with the soft limits first and collinear after.

\section{Energy--momentum tensor from shadow transform and double copy}\label{sec:shadow}
As discussed in the introduction, the set of BMS algebra generators consists of superrotations and supertranslations.
The energy momentum tensors $T(z), \bar{T}(\bz)$ encode the superrotation generators and the supertranslation field $P(z,\bz) $ encodes
the supertranslation generators. In this section we will follow an alternative approach to the Sugawara construction of the energy-momentum tensor.
We will follow an observation from \cite{Pate:2019lpp} to construct the energy-momentum tensor using a pair of dimension zero, opposite helicity gauge bosons. Inspired by this relation, we will propose a similar construction for the supertranslation field $P(z,\bz) $.

\subsection {A double copy construction of the energy momentum tensor}\label{sec:shadowKLT}

In the following we shall consider a pair of dimension zero gauge boson operators with
a shadow transform of one of the gauge bosons.
In general the shadow transform of an operator of the CCFT is given by the relation  \cite{Osborn:2012vt}:
\be\label{eq:shadow}
\widetilde{\cO^a_{\D, J}}(z,\bz)=\tilde\cO^a_{2-\Delta,-J}(z,\bz)={ K_{\D,J}\over \pi}  \int {d^2w \over (z-w)^{2-\D-J} (\bz-\bw)^{2-\D+J}} \ \cO^a_{\D,J}(w,\bw)\ .
\ee
where $K_{\D,J}=\D+J-1$\footnote{It seems there is a clash in the literature concerning the normalization factor for the shadow transform. In \cite{Dolan:2011dv} the normalization constant is $K_{\D,J}= {\Gamma(2-\D+J)\over \Gamma(\D+J-1)}$ unlike the one of \cite{donnay_conformally_2019,Pasterski:2017kqt} which we use in the main text. For $J=-1$ and as $\D\ra 0$, the normalization of \cite{Dolan:2011dv} behaves as $K_{\D,J}\sim \D$ and goes to zero. It is not clear why this discrepancy occurs, but in this case the only modification will be that in (\ref{Total}) we will need only the $(\D_1+\D_2)$ factor in the definition of the energy momentum tensor. The rest of our analysis leads though to the same conclusions. }.
Following (\ref{eq:sugawara_0}) we can introduce a modified energy momentum tensor on the celestial sphere
by choosing a pair of dimension zero gauge boson operators and considering the following expression
\begin{equation}\label{Start}
T(w_1)\sim \lim_{w_2\to w_1}\sum_a \delta^{ab}\ \cO^{a}_{0,+}(w_2,\bar w_2)\tilde\cO^b_{2,+}(w_1,\bar w_1)\ ,
\end{equation}
 based on (\ref{eq:sugawaranormal}) for level $k=0$.
 Here, the first gluon operator $\cO^{a}_{0,+}$ has spin one and vanishing dimension $\Delta_2=0$ (with $h_2=\tfrac{1}{2}+\tfrac{i\lambda_2}{2}, \bar h_2=-\tfrac{1}{2}+\tfrac{i\lambda_2}{2}$, $\lambda_2 \to 0$),
while the second  operator $\tilde\cO^b_{2,+}$ with $\tilde\Delta_1=2$ arises from a
shadow operation
\be\label{ShadowJ}
\tilde\cO^b_{2,+}(w_1,\bar w_1)\sim \int d^2z_1\ (z_1-w_1)^{-3}(\bar z_1-\bar w_1)^{-1}
\ \cO^b_{0,-}(z_1,\ov z_1)\ ,
\ee
of a gluon operator of negative spin one and vanishing dimension $\Delta_1=0$ (with $h_1=-\tfrac{1}{2}+\tfrac{i\lambda_1}{2}, \bar h_1=\tfrac{1}{2}+\tfrac{i\lambda_1}{2}$, $\lambda_1\to 0$). We have ignored the normalization factors of the shadow transform since they are not important for our arguments below and can be absorbed in an overall normalization for the energy momentum tensor.
To proceed we use the OPE of two gluon states  of opposite spins, which  can be found in \req{opemp}
\begin{align}
\cO^{a}_{\D_2,+}(w_2,\bar w_2)\cO^b_{\D_1,-}(z_1,\bar z_1)&=\fc{\Delta_1-1}{\Delta_2(\Delta_1+\Delta_2-2)}\sum_c\frac{\tilde{f}^{abc}}{w_2-z_1} \cO^{c}_{(\D_1+\D_2-1),-}(z_1,\bar z_1)
\nonumber \\
&-2 \delta^{ab}\ \frac{\bar w_2-\bar z_1}{w_2-z_1}\ \fc{(\Delta_1-1)(\Delta_1+1)(\Delta_2-1)}{\Delta_2(\Delta_1+\Delta_2)(\Delta_1+\Delta_2-1)}\ \cO_{(\Delta_1+\Delta_2),-2}(z_1,\bar z_1)\nonumber \\
&+ \tilde{f}^{abc}\ \Lambda(\D_1,\D_2)\ (\bw_2-\bz_1)\ \cO^c_{\D_1+\D_2+1,-}(z_1,\bz_1)+ \nonumber \\
&+\delta^{ab}\ M(\D_1,\D_2)\ (\bw_2-\bz_1)^2\ \cO_{\D_1+\D_2+2, -2}(z_1,\bz_1)+\ldots\ ,
%&+\sum_{k\geq 0} (w_2-z_1)^k\ {\cal C}_k(z_1,\ov z_1)\ .
\label{OPEgg}
\end{align}
where $\Lambda(\D_1,\D_2), M(\D_1,\D_2)$ are constants which depend on the details of the $D=4$  theory from which their OPE has been derived.
Above we have included possible single--pole or finite terms \cite{Pate:2019lpp}.
After inserting (\ref{ShadowJ}) into (\ref{Start}) and using \req{OPEgg}  we arrive at:
\begin{align}
&\lim_{\Delta_1,\Delta_2\rightarrow 0} \ \lim_{w_1\to w_2}
 \int d^2z_1\ (z_1-w_1)^{-3}(\bar z_1-\bar w_1)^{-1}\  \delta^{aa}\nonumber \\
&\times   \left\{ \fc{2}{\Delta_2(\Delta_1+\Delta_2)} \ \frac{\bar w_2-\bar z_1}{w_2-z_1}\ \cO_{\D_1+\D_2,-2}(z_1,\bar z_1)+  M(\D_1,\D_2) (\bw_2-\bz_1)^2\ \cO_{\D_1+\D_2+2, -2}(z_1,\bz_1)\right\}\label{Eq16}
%+\sum_{k\geq 0} (w_2-z_1)^k\ {\cal C}_k(z_1,\ov z_1)\right\}\label{Eq16}
\end{align}
Note, that the first and third  term of the OPE (\ref{OPEgg})  cancel after performing the color sum  in (\ref{Start}).
The first term of the equation above can be related to the energy--momentum tensor~\cite{Fotopoulos:2019tpe}:
\be
T(w)\sim\int d^2z\ (z-w)^{-4}\ \cO_{0,-2}(z,\bar z)\ .
\ee
% with the normalization $f(\Delta)=\h\tfrac{\Delta(\Delta-1)}{\Gamma(\Delta+2)}$.
In fact, after taking the limit $w_1\to w_2$ we obtain
\begin{align}
&2 \dim g\ \lim_{\Delta_1,\Delta_2\rightarrow 0}
\fc{1}{\Delta_2(\Delta_1+\Delta_2)}\ \int d^2z_1\ (z_1-w_1)^{-4}\ \cO_{0,-2}(z_1,\bar z_1)\nonumber\\
&=2 \dim g\ \lim_{\Delta_1,\Delta_2\rightarrow 0}
\fc{1}{\Delta_2(\Delta_1+\Delta_2)}\   T(w_1)\ ,
\end{align}
with the dimension $\dim g=\delta_{aa}$ of the underlying gauge group.
In total we have the following relation
\begin{equation}\label{Total}
T(w_1)=\fc{1}{\tilde{C}_2(G)}\ \lim_{\Delta_1,\Delta_2\rightarrow 0}[\Delta_2(\Delta_1+\Delta_2)]\lim_{w_2 \to w_1} \sum_a \ \cO^{a}_{\Delta_2,+}(w_2,\bar w_2)\tilde\cO^a_{2-\Delta_1,+}(w_1,\bar w_1),
\end{equation}
which assumes the desired form \req{Start}. The latter takes the Sugawara form (\ref{eq:sugawaranormal}) upon replacing the factor $\tfrac{1}{2\dim g}$ by $\tfrac{1}{2k+\tilde{C}_2(G)}$  for $k=0$.

Having fixed the normalization constant we can consider the regular terms of the OPE. The limit at $w_2\to w_1$ gives
\begin{align}\label{eq:regope_1}
\lim_{\Delta_1,\Delta_2\rightarrow 0}&[\Delta_2(\Delta_1+\Delta_2)]\ \delta^{aa} M(\D_1, \D_2)\nonumber \\
&\times \lim_{w_2 \to w_1}   \int d^2z_1\ (z_1-w_1)^{-3}(\bar z_1-\bar w_1)^{-1} (\bw_2-\bz_1)^2\ \cO_{\D_1+\D_2+2, -2}(z_1,\bz_1) \nonumber \\
&=\lim_{\Delta_1,\Delta_2\rightarrow 0}[\Delta_2(\Delta_1+\Delta_2)]\  \delta^{aa} \ M(\D_1, \D_2)  \int d^2z_1\ (z_1-w_1)^{-3} (\bz_1-\bw_1)\ \cO_{2, -2}(z_1,\bz_1).
\end{align}
%%  As explained in \cite{Pate:2019lpp} [cf. eq. (117)]
The action of this hard operator on primaries will lead to the following potentially singular terms
\begin{align} \label{eq:regope_2}
&\cO_{2, -2}(z_1,\bz_1)\cO^a_{\D, \pm}(w,\bw) \sim \rho(\D)\ {z_1-w\over \bz_1-\bw}\ \cO^a_{\D+2, \pm}(w,\bw)
\end{align}
Naively, after integration in (\ref{eq:regope_1}) we get, that close to the operator insertion the integral behaves as
\begin{align} \label{eq:regope_3}
\int d^2z_1\ { (\bz_1-\bw_1)\over (z_1-w_1)^{3}}\cO_{2, -2}(z_1,\bz_1)\cO^a_{\D, \pm}(w,\bw)\sim
%\sim \rho(\D)\int d^2z_1\ { (\bz_1-\bw_1)\over (z_1-w_1)^{3}} \frac{z_1-w}{\bar z_1-\bar w} \cO^a_{\D+2, \pm}(w,\bw)\sim\nonumber \\
 \rho(\D) {\bw_1-\bw\over w_1-w} \cO^a_{\D+2, \pm}(w,\bw)
\end{align}
where we have used standard conformal integrals (cf. \cite{Dolan:2011dv})
 \be\label{eq:confint}
 \int d^2z_1\ { (\bz_1-\bw_1)\over (z_1-w_1)^{3}} \frac{z_1-w}{\bar z_1-\bar w}=\pi\ \frac{\bar w-\bar w_1}{w-w_1}
 \ee
 to extract the singular part of this integral.
%{\bf where we used the following integral equation
%$$\int d^2z_1 \frac{z_1-w_1}{\bar z_1-\bar w_1} (z_1-w_1)^{-3}(\bar z_1-\bar w_1)^{-1}=\pi\
%\frac{\bar w-\bar w_1}{w-w_1}$$
%\be\label{eq:integral}	
%\int d^2z_1\ { (\bz_1-\bw_1)\over (z_1-w_1)^{3}} \frac{z_1-w}{\bar z_1-\bar w}=\pi\
%\frac{\bar w-\bar w_1}{w-w_1}
%\ee
%This integral can be computed using the conformal integrals  of \cite{Dolan:2011dv}
%\be\label{eq:confintegral}
%I_2= {1\over 2\pi}\int d^2z\ \prod\limits_{i=1}^2 {1\over (z-z_i)^{h_i}} {1\over (\bz-\bz_i)^{\bh_i}}={\Gamma(1-h_1)\Gamma(1-h_2)\over \Gamma(\bh_1)\Gamma(\bh_2)}(-1)^{h_1-h_2} 2\pi \delta^{(2)}(z_1-z_2)
%\ee
%with $\sum_{i=1}^2 h_i=\sum_{i=1}^2\bh_i=2, \ h_i-\bh_i \in \mathbb{Z}$. Indeed we can verify that (\ref{eq:integral})
%\be\label{eq:confintegral2}
%\bpartial_{\bw}^2 \int d^2z_1\ { (\bz_1-\bw_1)\over (z_1-w_1)^{3}} \frac{z_1-w}{\bar z_1-\bar w}=2I_2= 2 \pi^2 \delta^{(2)}(w_1-w)
%\ee}
Equation (\ref{eq:regope_3}) leads to singular behaviour, either pole type if we consider separately the holomorphic limit  $w\to w_1$ or of a singular angular distribution if both $w\to w_1$ and $\bw\to \bw_1$. If we consider strictly EYM theory this term does not exist. As explained in \cite{Pate:2019lpp}, the subleading term in the OPE (\ref{OPEgg}) originates from higher derivative bulk interactions of the form  $R F^2$. So in the pure EYM case, these are absent and the final result is given by (\ref{Total}) and we have demonstrated the desired result.

Nevertheless,  we are interested in the energy-momentum tensor for more general theories with higher derivative corrections i.e. $R F^2$ due to quantum, stringy or other effects. The important point is that such corrections will not contribute to the proposal (\ref{Start}).  We will demonstrate, that the constant $M(\D_1,\D_2)$ has at most single poles under the double soft limit $\D_1,\D_2 \to0$. Then  in (\ref{eq:regope_1})  the last term will drop in the double soft limit $\D_1,\D_2\to 0$. The prefactor $\D_2(\D_1+\D_2)$ goes to zero quadratically but $M(\D_1,\D_2)$ has a single pole.
% So the term eventually will drop when the operator is inserted in a correlator.
To see this we need to follow the discussion of Appendix A in \cite{Pate:2019lpp}. The term of interest in the OPE (\ref{OPEgg}) stems from $RF^2$ higher derivative corrections of EYM .

In general the cubic vertex has the form
\begin{align}\label{eq:cubvert}
 V=\partial^m \Phi_1(x) \Phi_2(x) \Phi_k(x)
 \end{align}
where the fields $\Phi$ can be $A_\mu, h_{\mu \nu}$ but Lorentz indices are suppressed and
the total number of derivatives $m$ distributed among all three fields $\Phi_1. \Phi_2, \Phi_k$.  The net dimension of the vertex is $d_V= 3+m$.
A Mellin transform analysis  of the collinear limit in a celestial amplitude, leads to the following result \cite{Pate:2019lpp}
\begin{align}\label{eq:mellincub}
\cA\sim \sum_{\a, \b} B(\D_1+m+\a-1,\D_2+\b-1)\int\limits_0^\infty d\omega_P\  \omega_P^{\D_1+\D_2+m-3} A_{\a,\b}(z_1,\bz_1, z_2,\bz_2, \omega_P,\dots),
\end{align}
where $\omega_P=\omega_1+\omega_2$ and in our case  $m=4$. In this case the operator $\Phi_k$ has dimensions $\D_k= \D_1+\D_2 +m-2\rightarrow 2$ which is the dimension of $\cO_{2,-2}$ in (\ref{eq:regope_1}). The remaining Mellin transform is a celestial amplitude with a hard operator insertion $\cO_{2,-2}$ and no poles are expected unlike for soft operators with dimension one. The labels $\a,\b$ determine the different powers of the energy factors in the collinear splitting functions \cite{fan_soft_2019,Pate:2019lpp,Fotopoulos:2019vac}
\be\label{eq:split}
Spit_{s_1,s_2}^s(p_1,p_2)= {1\over z_{12}}{ \omega_1^{m+\a} \omega_2^\b \over \omega_P^{\a+\b}} {1\over \omega_1 \omega_2}\ ,
\ee
with $s_i$ the helicities of the collinear states and $\a,\b\geq -1$ in YM and $\a,\b\geq -2$  in GR.
In higher derivative theories with couplings  $RF^2$ etc. they are always $\a,\b \geq -m$.

As it is clear from the derivation of (\ref{eq:mellincub})
\be\label{eq:M}
M(\D_1,\D_2)= c\ B(\D_1+m+\a-1,\D_2+\b-1)\ ,
\ee
where $c$ is a numerical constant independent of the dimensions $\D_1,\D_2$.
We notice that in the limit $\D_1 , \D_2\to 0$, at most single poles can appear in the prefactor $ B(\D_1+m+\a-1,\D_1+\b-1)\to  B(3+\a,\b-1)$.  Actually, our result is more general, since the Beta function has at most single poles and could be applied to arbitrary higher derivative corrections.

%These are single poles if $m+\a-1= 3+\a<0$. The OPE produces a operator with dimension $\D_k= \D_i+\D+ 2\to \D+2$ and cannot give additional poles as $\D_i\to 0$.
This concludes the proof that the prefactor $M(\D_1,\D_2)$ in (\ref{eq:regope_1}) has at most a single pole as $\D_1,\D_2 \to 0$. In (\ref{eq:regope_1}) we see that automatically the limit leads to zero since we have a double zero from the overall prefactor. We conclude that the result (\ref{Total}) holds for more general extensions of EYM.

The Sugawara inspired relation (\ref{Total}) gives a gauge gravity relation established as a
relation between a pair of gauge boson operators
$\cO^{a}_{0,+}, \cO^b_{0,-}$ and a graviton operator $\cO_{0,-2}$ on the celestial sphere. Notice that this is not the usual Sugawara construction since the operators $\cO^{a}_{0,\pm}$ are not dimension one and do not generate a Kac-Moody symmetry. Having said this, it seems that our construction is more like the KLT or double--copy equivalent in CCFT.
Note, that the well-known KLT relations express gravitational amplitudes as sums over squares of gauge amplitudes supplemented by a momentum dependent kernel. The latter accounts for  disentangling monodromy relations on the
string world--sheet. On the other hand, (\ref{Total}) gives a direct relation between a graviton and a pair of gauge bosons on the celestial sphere without any additional momentum dependent factors.

\subsection{A double copy construction for supertranslations}\label{sec:KLTP}

In \cite{donnay_conformally_2019,Fotopoulos:2019vac}, it was shown that supertranslations are generated by the operator $P$
\be\label{pdef} P(z,\bar z)\equiv\partial_{\bz} {\cal O}_{\D\to 1,+2}\ ,\ee
where its OPE with  spin one primaries is given by the relation:
\be\label{poope1}
P(z,\bz)\cO_{\Delta,J}(w,\bw)=\frac{(\D-1)(\D+1)}{4\D} \frac{1}{z-w}\cO_{\Delta+1,J}(w,\bw) +\makebox{regular}\ ,\ J=\pm 1\ .
\ee
The presence of $(\D-1)$ factors in the above OPE coefficients implies that the products $P(z)j^a(w)$, $P(z)\bar j^a(\bw)$ are regular. Similar relations hold for the antiholomorphic operator $\bar{P}$. An important property is that supertranslations shift the dimension of the usual fields up $\D \to \D+1$ or equivalently $(h,\bh)\ra (h+\ha,\bh+\ha)$. This may be checked  by applying in particular the momentum operator which generates translations along the light-cone direction (see \cite{Stieberger:2018onx,Fotopoulos:2019vac}):
\be \label{eq:P00}
P_{-\ha, -\ha}=P_0+P_3=e^{(\partial_h+ \partial_{\bar{h}})/2}\ .
\ee

Proceeding one step ahead one can try a kind of double copy construction of the operator $P(z,\bz)$ of (\ref{pdef}) as well. We will show that:
\be\label{eq:Psugawara}
 {\cal O}_{\D\to 1,+2}(w_1,\bw_1)\sim  \lim_{\Delta_2 \to 0 ,\Delta_1 \to 1}\lim_{w_2 \to w_1} \D_2 \sum_a \cO^a_{\D_2, +}(w_2,\bw_2) \tilde{\cO}^a_{2-\D_1,+} (w_1,\bw_1)\ .
\ee
To prove the equivalence above, we use once more the OPE (\ref{OPEgg}).
 \begin{align}  \label{eq:ggOPEb}
&\lim_{\Delta_1\ra 1,\Delta_2\rightarrow 0} \ \lim_{w_1\to w_2}
 \int d^2z_1\ (z_1-w_1)^{-2}\   \\
&\times  \delta^{aa} \left\{ \fc{4}{\Delta_2} \ \frac{\bar w_2-\bar z_1}{w_2-z_1}\ \cO_{\D_1+\D_2,-2}(z_1,\bar z_1)
+  M(\D_1,\D_2)\ (\bw_2-\bz_1)^2\ \cO_{\D_1+\D_2+2, -2}(z_1,\bz_1)\right\}\ . \nonumber
%+\sum_{k\geq 0} (w_2-z_1)^k\ {\cal C}_k(z_1,\ov z_1)\right\}\label{Eq16}
\end{align}
In the equation above we need to chose a specific order of limits $\D_2\to 0$ first and $\D_1\to 1$ last.
The first term is the one we need for our purpose.  For EYM there are no higher derivative terms and the leading term is all we need.
As in the case of the energy-momentum tensor, for general theories beyond EYM,  we need to analyze the potential implications of the subleading operator $ \cO_{\D_1+\D_2+2, -2}\to  \cO_{3,-2}$.  The  coefficient $M(\D_1,\D_2)$ of the subleading term can have a single pole as $\D_2\to 0$ following the collinear limits of celestial amplitudes (\ref{eq:mellincub}). So naively this term can create additional contributions in the proposal (\ref{eq:Psugawara}). Nevertheless, this operator has the following OPEs with primary operators:
\begin{align} \label{eq:regope_2b}
&\cO_{3, -2}(z_1,\bz_1)\cO^a_{\D, \pm}(w,\bw) \sim \rho(\D)\ {z_1-w\over \bz_1-\bw}\ \cO^a_{\D+3, \pm}(w,\bw)\ .
\end{align}
Applying the integration in  (\ref{eq:ggOPEb}) the integrant near the operator insertion behaves as follows
\begin{align} \label{eq:regope_3b}
\int d^2z_1\ & (z_1-w_1)^{-2} (\bz_1-\bw_1)^2\  \cO_{3, -2}(z_1,\bz_1)\cO^a_{\D, \pm}(w,\bw)\nonumber\\
&\sim \rho(\D) \int d^2z_1\ { (\bz_1-\bw_1)^2 \over (z_1-w_1)^{2}} {z_1-w \over \bz_1-\bw }\  \cO^a_{\D+3, \pm}(w,\bw)\ .
\end{align}
Unlike (\ref{eq:regope_3}) no poles can emerge from this expression. So finally, we derive
\ba\label{eq:relation3}
 &&\lim_{\Delta_2 \to 0 ,\Delta_1 \to 1}\lim_{w_2 \to w_1} \D_2 \sum_a \cO^a_{\D_2, +}(w_2,\bw_2) \tilde{\cO}^a_{2-\D_1,+} (w_1,\bw_1)\nonumber \\
 &&= \lim_{\Delta_2 \to 0 ,\Delta_1 \to 1}\lim_{w_2 \to w_1} \int {d^2z \over (w_1-z)^2} \ {\bw_2-\bz \over w_2-z} \cO_{\D_1+\D_2, -2}(z,\bz)\nonumber\\
 &&=  \int d^2z   \ {\bw_1-\bz \over (w_1-z)^3}\ \cO_{1, -2}(z,\bz) \nonumber\\
 && =\widetilde{{\cO}_{1,-2}}(w_1,\bw_1)= \tilde{\cO}_{1,+2}(w_1,\bw_1)=\cO_{1,+2}(w_1,\bw_1)\ ,
\ea
%where we used similar argument as in section \ref{sec:shadowKLT}  to show that the subleading terms of the OPE of the two conformally soft operators do not produce additional singular terms.
where in the last step we have used the relation for dimension one operators  $\tilde{\mathcal{O}}^a_{1, +} (w_1)= \mathcal{O}^a_{1, +} (w_1)$ \cite{Pasterski:2017kqt}. For completeness we give the OPE of the operators $ \cO_{1,+2}(w)$ with spin one primaries $J=\pm 1$,
\be\label{eq:relation 2}
{\cal O}_{\D\to 1,+2}(z,\bz) \cO_{\D_i,J}(w,\bar w)\sim \frac{(\D-1)(\D+1)}{4\D}\frac{\bar{z}-\bar{w}}{z-w}\cO_{\D_{i}+1,J}(w,\bw)
\ee
from \cite{Fotopoulos:2019vac}. Then applying this on (\ref{pdef}) we derive the OPE of $P(z,\bz)$ with primaries
\be\label{eq:relation 3}
P(z,\bz) \cO_{\D_i, J}(w,\bar w)\sim\frac{(\D-1)(\D+1)}{4\D} \frac{1}{z-w}\cO_{\D_{i}+1,+}(w,\bw)
\ee
One mode of this field is the operator $P_{-\ha,-\ha}$ in (\ref{eq:P00}).

	\section{Conclusions}
	
From the  study of scattering amplitudes in four--dimensional Minkowski space--time
some striking relations between gravity and gauge amplitudes have emerged.
For a review see \cite{Bern:2019prr}. These observations suggest a deeper connection between  gauge and gravity theories and indicate  the existence of some gauge structure in quantum gravity. However, the origin  of  these relations is  yet poorly understood in four--dimensional Minkowski space--time.
The Mellin transform of gauge and gravitational states and amplitudes to celestial sphere gives a new way of looking at
quantum field theory and quantum gravity and might shed light on the underlying symmetries of these amplitude relations.
In particular, it seems feasible  that the manifestation of  double--copy--constructions may have a simpler emergence when considered within the underlying conformal field theory on the celestial sphere.

In this work we discussed the energy-momentum tensor of the pure gauge sector of the CCFT. 	 For the pure gauge theory, it has been suggested \cite{He:2015zea, Cheung:2016iub}, that a particular subsector of the CCFT, the one of soft operators, can be described by a current algebra, a Kac-Moody algebra. In this work, we used the Sugawara method to construct the energy-momentum tensor $T^S(z)$ from the celestial amplitude of gluons. From the analysis of the soft and collinear limits of gluon amplitudes we extracted the OPEs of the Sugawara energy momentum tensor with primary fields of the CCFT. The OPE of the holomorphic Sugawara energy-momentum tensor has the expected form for soft holomorphic operators $j^a(z)$ which correspond to soft positive helicity gluons. For antiholomorphic soft operators and hard operators, the OPE is not as expected. A modification will be necessary. We discussed these shortcomings and suggested potential resolutions on how to decouple the sub-CFT that describes the positive helicity soft sector from the rest of the theory. We also developed several gauge group identities, some of which are novel and potentially useful for scattering amplitude computations in general.

Subsequently we used CCFT OPEs for EYM theory to construct from a pair of gluon operators the energy momentum tensor and the supertranslation operator of the BMS algebra. This method bears resemblance to the double-copy method  that relates gauge and gravity amplitudes. The energy momentum tensor we constructed has the correct action on both the soft and hard operators of the theory. It is a generalization of the Sugawara method, although the Kac-Moody current algebra origin of this construction is not so clear.

There are several open questions which deserve further study. In section \ref{sec:comments} we discussed the importance of massive states in relation to the soft sub-sector of the theory. Massive states should correspond to Wilson lines on the CCFT and it is an interesting question how to implement them in the celestial amplitudes picture. It is important to investigate correlators of soft operators with Wilson lines and extract the OPE with the Sugawara energy momentum tensor. Finally, the BMS algebra on the CCFT language was discussed recently in \cite{Fotopoulos:2019vac}. It would be interesting to compute the algebra using the Sugawara energy-momentum tensor and see if we can have a BMS type of symmetry for the soft sub-sector of the theory.
\\

\vskip 1mm
\leftline{\noindent{\bf Acknowledgments}}
\vskip 1mm
\noindent
We are grateful to Bin Zhu for collaboration in related topics and correspondence. StSt and TT are grateful to Monica Pate, Ana-Maria Raclariu and Andy Strominger for useful conversations.
This material is based in part upon work supported by the National Science Foundation
under Grant Number PHY--1913328.
Any opinions, findings, and conclusions or recommendations
expressed in this material are those of the authors and do not necessarily
reflect the views of the National Science Foundation.	
	
	%\renewcommand{\thesection}{A}
	%\setcounter{equation}{0}
	%\renewcommand{\theequation}{A.\arabic{equation}}
	%\renewcommand{\thesection}{A}
	%\setcounter{equation}{0}
	%\renewcommand{\theequation}{A.\arabic{equation}}
	%\vskip 3mm
	%\begin{center}{\noindent{\large\bf Appendix}}\end{center}\noindent

\appendix      %JHEP provides us with this environment of Appendix	

	\section{Solution of the $\mathbf{n}$-particle momentum-conservating delta function}
	\label{sec:appendixA}

	We use \cite{fan_soft_2019} and give an expression of the $n$-particle momentum-conservating delta functions which appear in the amplitudes (\ref{eq:mellin_partial_amp}) in terms of energies $\omega_i$ and celestial coordinates $z_i,\bz_i$. For the  $n$-particle ($n\ge 5)$  momentum-conservating delta function, we choose to use the first four energies $\omega_1, \omega_2, \omega_3,\omega_4$ to localize the solution. This choice is arbitrary and any other choice also works.
	Define the following cross-ratios of celestial coordinates:
	\begin{equation}
	\label{eq:cross_ratio}
	t_i = \frac{z_{12}z_{3i}}{z_{13}z_{2i}}, \quad i=4,5,\ldots, n.
	\end{equation}
	Then the $n$-point momentum delta function is solved as
	\begin{equation}
	\label{eq:delta_solution}
	\delta^4\big(\sum_{i=1}^{N} \epsilon_i \omega_i q_i\big) = \frac{i}{4}\frac{(1-t_4)(1-\bar{t}_4)}{t_4-\bar{t}_4} \frac{1}{|z_{14}|^2|z_{23}|^2} \prod_{i=1}^{4}\delta(\omega_i-\omega_i^\star).
	\end{equation}
	The solutions for the  four chosen energies are
	\begin{equation}
	\label{eq:energy_sol}
	 \omega_i^\star=f_{i5}\omega_5+f_{i6}\omega_6+\ldots+f_{in}\omega_n\ ,
	\end{equation}
	where $f_{ij}, i=1,2,3,4, j=5,6,\ldots,n$ are   functions of cross-ratios:
	\begin{align*}
		f_{1j}&=t_4 \Big|\frac{z_{24}}{z_{12}}\Big|^2 \frac{(1-t_4)(1-\bar{t}_4)}{t_4-\bar{t}_4}\epsilon_1\epsilon_j\frac{t_j-
	\bar{t}_j}{(1-t_j)(1-\bar{t}_j)}\Big|\frac{z_{1j}}{z_{14}}\Big|^2   - \epsilon_1\epsilon_j t_j\Big|\frac{z_{2j}}{z_{12}}\Big|^2\ , \\
		f_{2j}&= - \frac{1-t_4}{t_4} \Big|\frac{z_{34}}{z_{23}}\Big|^2 \frac{(1-t_4)(1-\bar{t}_4)}{t_4-\bar{t}_4}\frac{\epsilon_1\epsilon_j}{\epsilon_1
	\epsilon_2}\frac{t_j-\bar{t}_j}{(1-t_j)(1-\bar{t}_j)}\Big|\frac{z_{1j}}{z_{14}}\Big|^2 + \frac{\epsilon_1\epsilon_j}{\epsilon_1\epsilon_2} \frac{1-t_j}{t_j}\Big|\frac{z_{3j}}{z_{23}}\Big|^2\ ,\\
		f_{3j}&= (1-t_4)\Big|\frac{z_{24}}{z_{23}}\Big|^2 \frac{(1-t_4)(1-\bar{t}_4)}{t_4-\bar{t}_4}\frac{\epsilon_1\epsilon_j}{\epsilon_1
	\epsilon_3}\frac{t_j-\bar{t}_j}{(1-t_j)(1-\bar{t}_j)}\Big|\frac{z_{1j}}{z_{14}}\Big|^2 - \frac{\epsilon_1\epsilon_j}{\epsilon_1\epsilon_3} (1-t_j)\Big|\frac{z_{2j}}{z_{23}}\Big|^2\ , \\
		f_{4j}&= - \frac{(1-t_4)(1-\bar{t}_4)}{t_4-\bar{t}_4}\frac{\epsilon_1\epsilon_j}{\epsilon_1
	\epsilon_4}\frac{t_j-\bar{t}_j}{(1-t_j)(1-\bar{t}_j)}\Big|\frac{z_{1j}}{z_{14}}\Big|^2.
	\end{align*}

	\section{Seven--gluon NMHV amplitude and $\mathbf{T^S\bar{j}}$ OPE}
	\label{sec:appendixB}

	In this appendix we will compute  the mixed $T^S(z)\bar{j}(\bar{w})$-OPE in the seven--gluon NMHV amplitude $A^{7}(-,-,-,+,+,+,+)$. We use the last two operators to define the Sugawara energy momentum tensor $T^S(z_{7})\sim\lim_{z_{6}\to z_{7}} j^a(z_{6})j^a(z_{7})$, and extract its OPE with the first operator $T^S(z_{7})\bar{j}(\bar{z}_1)$.
	
	The explicit form  of the subamplitude $A^{7}(-,-,-,+,+,+,+)$ was obtained by the BCFW method in reference~\cite{Britto:2004ap}, which reads as:
	\begin{equation}
	\label{eq:7nmhv_1}
	\begin{array}{l}
	A\left(1^{-}, 2^{-}, 3^{-}, 4^{+}, 5^{+}, 6^{+}, 7^{+}\right)= 	\frac{\langle 1|2+3| 4]^{3}}{t_{2}^{[3]}\langle 56\rangle\langle 67\rangle\langle 71\rangle[23][34]\langle 5|4+3| 2]} \\
	-\frac{1}{\langle 34\rangle\langle 45\rangle\langle 6|7+1| 2]}\left(\frac{\langle 3|(4+5)(6+7)| 1\rangle^{3}}{t_{3}^{[3]} t_{6}^{[3]}\langle 67\rangle\langle 71\rangle\langle 5|4+3| 2]}+\frac{\langle 3|2+1| 7]^{3}}{t_{7}^{[3]}\langle 65\rangle[71][12]}\right)\ .
	\end{array}
	\end{equation}
	Taking limit of $z_6\to z_7$ for defining the Sugawara $T(z_7)$ and the limit of $z_7\to z_1$ for extracting the mixed OPE, this seven--gluon NMHV amplitude has the following leading order poles in terms of celestial coordinates
	\begin{align}
	\label{eq:7nmhv_ope_1}
	\lim_{z_6\to z_7\to z_1}A^{7}&(---++++)=-
	\frac{\omega_{1}\omega_4\omega_5}{\omega_2\omega_3\omega_6\omega_7(\omega_1+\omega_6 +\omega_{7})^2} \frac{\bar{z}_{45}^3}{\bar{z}_{12}\bar{z}_{23}\bar{z}_{34}\bar{z}_{51}}(\frac{1}{ z_{67}z_{71}}) \nonumber\\
	&-
	\frac{\omega_{2}\omega_3(\omega_6 +\omega_{7})^2}{\omega_1\omega_4\omega_5\omega_6\omega_7(\omega_1+\omega_6 +\omega_{7})^2} \frac{z_{23}^3}{z_{12}z_{34}z_{45}z_{51}}(\frac{1}{ z_{67}\bar{z}_{71}}) \nonumber\\
	&-	\frac{\omega_{2}\omega_3\omega_7 }{\omega_1^2(\omega_1+\omega_{7})\omega_4 \omega_5\omega_6} \frac{z_{23}^3}{z_{12}z_{34}z_{45}z_{51}}(\frac{1}{ z_{71}\bar{z}_{71}}).
	\end{align}
	The Mellin integral of the first term in the above equation is
	\begin{align}
	\label{eq:7nmhv_int_0}
	&g(\lambda_1)g(\lambda_6)g(\lambda_{7})\int_0^\infty d\omega_1 d\omega_6 d\omega_7\,  \omega_1^{i\lambda_1}\omega_6^{i\lambda_6}\omega_7^{i\lambda_7} \frac{\omega_1}{ \omega_6\omega_7 (\omega_1+\omega_6 +\omega_7)^2}\ldots \nonumber\\
	&=g(\lambda_1)g(\lambda_6)g(\lambda_7)\int_0^\infty d\omega'_1\int_0^{\omega'_1} d\omega'_6 \int_0^{\omega'_6} d\omega_{6}\, \omega_6^{-1+i\lambda_6} (\omega'_6 - \omega_6)^{-1+i\lambda_7}(\omega'_1 - \omega'_6)^{1+i\lambda_{1}}\omega\prime_1^{-2}\ldots  \nonumber\\
	&=g(\lambda_1)g(\lambda_6)g(\lambda_7)B(i\lambda_6,i\lambda_7)B(i\lambda'_6,2+i\lambda_{1})\int_0^\infty d\omega'_1 {\omega'}_1^{-1+i\lambda'_1} \ldots,
	\end{align}
	where the integral is performed with change of variable $\omega'_6=\omega_6+\omega_7$, $\omega'_1=\omega_1+\omega'_6$
	and we have defined new quantities $\lambda'_6=\lambda_6+\lambda_7$, $\lambda'_1=\lambda_1+\lambda'_6$.
	In the conformal soft limit  $ \lambda_{6},\lambda_7 \to 0$, the integral is nonzero
	\begin{align}
	\label{eq:7nmhv_int_1}
	&g(\lambda_1)g(\lambda_6)g(\lambda_7)B(i\lambda_6,i\lambda_7)B(i\lambda'_6,2+i\lambda_{1})\int_0^\infty d\omega'_1 {\omega'}_1^{-1+i\lambda'_1} \ldots \nonumber\\
	&= \frac{i\lambda_6\Gamma(i\lambda_6)}{\Gamma(2+i\lambda_6)} \frac{i\lambda_7 \Gamma(i\lambda_7)}{\Gamma(2+i\lambda_7)} \frac{i\lambda_{1}}{\Gamma(2+i\lambda'_{1})} \int_0^\infty d\omega'_1 {\omega'}_1^{1+i\lambda'_1} \ldots =g(\lambda_{1}) \int_0^\infty d\omega_1 {\omega}_1^{1+i\lambda_1} \ldots,
	\end{align}
	where in the last step we have relabelled $\omega'_1$ as $\omega_1$.
	Similarly, the Mellin integral of the second term in eq.~\eqref{eq:7nmhv_ope_1} is
	\begin{align}
	\label{eq:7nmhv_int_2}
	&g(\lambda_1)g(\lambda_6)g(\lambda_{7})\int_0^\infty d\omega_1 d\omega_6 d\omega_7\,  \omega_1^{i\lambda_1}\omega_6^{i\lambda_6}\omega_7^{i\lambda_7} \frac{ (\omega_6 +\omega_7)^2}{ \omega_1\omega_6\omega_7 (\omega_1+\omega_6 +\omega_7)^2}\ldots \nonumber\\
	&=g(\lambda_1)g(\lambda_6)g(\lambda_7)\int_0^\infty d\omega'_1\int_0^{\omega'_1} d\omega'_6 \int_0^{\omega'_6} d\omega_{6}\, \omega_6^{-1+i\lambda_6} (\omega'_6 - \omega_6)^{-1+i\lambda_7}(\omega'_1 - \omega'_6)^{-1+i\lambda_{1}}\omega\prime_6^2 \omega\prime_1^{-2}\ldots  \nonumber\\
	&=g(\lambda_1)g(\lambda_6)g(\lambda_7)B(i\lambda_6,i\lambda_7)B(2+i\lambda'_6,i\lambda_{1})\int_0^\infty d\omega'_1 {\omega'}_1^{-1+i\lambda'_1} \ldots.
	\end{align}
	In the conformal soft limit  $ \lambda_{6},\lambda_7 \to 0$, the integral is zero
	\begin{align}
	\label{eq:7nmhv_int_3}
	&g(\lambda_1)g(\lambda_6)g(\lambda_7)B(i\lambda_6,i\lambda_7)B(2+i\lambda'_6,i\lambda_{1})\int_0^\infty d\omega'_1 {\omega'}_1^{-1+i\lambda'_1} \ldots \nonumber\\
	&= i\lambda'_6(1+i\lambda'_6) \frac{i\lambda_6\Gamma(i\lambda_6)}{\Gamma(2+i\lambda_6)} \frac{i\lambda_7 \Gamma(i\lambda_7)}{\Gamma(2+i\lambda_7)}  \frac{i\lambda_{1} \Gamma(i\lambda_{1})}{\Gamma(2+i\lambda_{1})\Gamma(2+i\lambda'_{1})} \int_0^\infty d\omega'_1 {\omega'}_1^{-1+i\lambda'_1} \ldots \nonumber\\
	&=0.
	\end{align}
Finally, the Mellin integral of the third term in eq.~\eqref{eq:7nmhv_ope_1} is
	\begin{align}
	\label{eq:7nmhv_int_4}
	&g(\lambda_1)g(\lambda_6)g(\lambda_{7})\int_0^\infty d\omega_1 d\omega_6 d\omega_7\,  \omega_1^{i\lambda_1}\omega_6^{i\lambda_6}\omega_7^{i\lambda_7} \frac{ \omega_7}{ \omega_1^2\omega_6 (\omega_1+\omega_7)}\ldots \nonumber\\
	&=g(\lambda_1)g(\lambda_7)\int_0^\infty d\omega_1  d\omega_7\,  \omega_7^{1+i\lambda_7} \omega_1^{-2+i\lambda_1}(\omega_1+ \omega_7)^{-1}\ldots  \nonumber\\
	&=g(\lambda_1)g(\lambda_7)B(-1+i\lambda_1,2+i\lambda_7)\int_0^\infty d\omega'_1 {\omega'}_1^{-1+i(\lambda_1+\lambda_7)} \ldots,
	\end{align}
	where in the first line we took the conformal soft limit of $\lambda_6\to 0$ and used the following formula~\cite{Nandan:2019jas}
	\begin{equation}
	\lim_{\lambda_6=0}g(i\lambda_6)\int d\omega_6 \omega_6^{-1+i\lambda_6} = \lim_{\lambda_6=0}\int d\omega_6 i\lambda_{6}\omega_6^{-1+i\lambda_6}=1.
	\end{equation}
	In the conformal soft limit  $\lambda_7 \to 0$, the integral is zero
	\begin{align}
	\label{eq:7nmhv_int_5}
	&g(\lambda_1)g(\lambda_7)B(-1+i\lambda_1,2+i\lambda_7)
	= \frac{i\lambda_7\Gamma(2+i\lambda_7)}{\Gamma(2+i\lambda_7)}g(\lambda_1) \frac{\Gamma(-1+i\lambda_1)}{\Gamma(1++i\lambda_1+i\lambda_7)}
	&\overset{\lambda_7=0}{=}0.
	\end{align}
	
	Combining the above three Mellin integrals, the final result is
	\begin{align}
	\label{eq:7nmhv_ope_nontrivial}
	\mathcal{A}^{7}(---++++)\overset{z_7\to z_6\to z_1}{=} -\lf(\frac{1}{z_{67}z_{71}}\ri) \mathcal{A}^5(1^-2^-3^-4^+5^+).
	\end{align}
	Adding the contribution for $z_6\leftrightarrow z_7$ we derive
	\begin{align}
	\label{eq:7nmhv_ope_nontrivial2}
	\mathcal{A}^{7}(---++++)\overset{z_7, z_6\to z_1}{=} -\lf(\frac{1}{z^2_{71}}\ri) \mathcal{A}^5(1^-2^-3^-4^+5^+).
	\end{align}
	Obviously $T(z_7)\bar{j}_(\bar{z}_1)$-OPE  has a double pole and it is not zero.

\section{Sugawara OPE with soft shadow operators}
	\label{sec:appendixC}	
	
In this appendix we discuss the role of the Sugawara energy momentum tensor in correlators with insertions of the soft ($\D\to1$) shadow of spin one conformal primary operators. The shadow of an operator is given by (\ref{eq:shadow}). For the case of spin one and dimension one operator this becomes
\be\label{eq:shadowD=1}
\widetilde{j}^a(z)=-{1\over 2\pi}\int {d^2w \over (z-w)^2} \bar{j}^a(w)
\ee
In \cite{Pasterski:2017kqt} the conformal primary wave functions for dimension one operators were shown to be equivalent to their shadow transforms. Nevertheless, in \cite{donnay_conformally_2019} it was shown that at the subleading order of the $\D\to1$ limit the two conformal primary wave functions differ by a logarithmic mode. We will leave the operator $\widetilde{j}^a_+(z)$ as a distinct operator from $j^a(z)$.

In order to discuss the algebra of the shadow  currents with the holomorphic currents $j^a(z)$, we will need the conformal  soft theorems (or OPEs)  of both $j^a(z)$ and $\bar{j}^a(\bz)$. At this point  it is important to distinguish two different situations depending on the order of the  consecutive soft limits. This is important in the case of opposite helicity gluons only. The OPEs in (\ref{eq:jbjope})  correspond to the case where positive helicity gluons are taken soft before negative ones and vice versa for  (\ref{eq:bjjope}). The action of the  shadow currents on hard primary operators has no ambiguity and agrees with (\ref{eq:JOope},\ref{eq:JOmope}).

Let us discuss case 1. Using the OPEs in (\ref{eq:jbjope}) we derive
\ba\label{eq:jshjalg1}
&& j^a(z) \bar{j}^{b}(\bw)\sim {f^{a b c}\  \bar{j}^c(\bw) \over z-w},  \quad  \widetilde{j}^a(z) \bar{j}^{b}(\bw)\sim {f^{a b c}\  \bar{j}^c(\bw) \over z-w}\\
&&j^a(z) j^{b}(w)\sim {f^{a b c}\  j^c(w) \over z-w} , \quad  j^a(z)\widetilde{j}^{b}(w)\sim reg, \quad    \widetilde{j}^a(z)\widetilde{j}^{b}(w)\sim reg \nonumber
\ea
%\ba\label{eq:jshjalg1}
%&& j^a(z) \bar{j}^{b}(\bw)\sim {f^{a b c}\  \bar{j}^c(\bw) \over z-w},  \quad  \widetilde{j}^a(z) \bar{j}^{b}(\bw)\sim {f^{a b c}\  \bar{j}^c(\bw) \over z-w}\\
%&&j^a(z) j^{b}(\bw)\sim {f^{a b c}\  j^c(\bw) \over z-w} , \quad  j^a(z)\widetilde{j}^{b}(\bw)\sim reg, \quad    \widetilde{j}^a(z)\widetilde{j}^{b}(\bw)\sim reg \nonumber
%\ea
%\mycolor{red}{Maybe the above equation should be (three typos of bar are changed)
%
%}
Since we are discussing the OPE of the Sugawara tensor with a soft operator we can use the method of section \ref{sec:gengroup}.  It is straight forward to apply the derivation there and derive the following OPE
\be\label{eq:Tsj1ope}
T^S(z) \widetilde{j}^a(w)\sim reg
\ee
This implies that the shadow currents $\widetilde{j}^a$ are inert under the conformal transformations generated by the Sugawara energy momentum tensor. This again leads to the necessity to modify the energy momentum tensor to account for the conformal transformation properties of the $\widetilde{j}^a(z)$ holomorphic currents. Nevertheless, this is more promising than considering correlators with antiholomorphic currents, since we found that contrary to expectations the holomorphic  Sugawara energy momentum tensor acts on those currents which have weights $(0,1)$ and should be normally inert. Of course the result above does not apply to the case of MHV amplitudes, since conformal soft limit of negative helicity gluons leads to a vanishing correlator. For the case of $N^kMHV$ though Mellin plus shadow transform lead to
\be\label{eq:NkMHVcorr}
A_n(g_1^-, g_2^-, \ldots g_ k^-, g_{k+1}^+,\ldots g_n^+) \longrightarrow \langle \widetilde{j}(z_1)\ldots  \widetilde{j}(z_k) j(z_{k+1}) \ldots j(z_n)\rangle
\ee
which is non vanishing generally and purely holomorphic.
\\
For case 2, similarly we derive
\ba\label{eq:jshjalg2}
&&j^a(z) j^{b}(w)\sim {f^{a b c}\  j^c(w) \over z-w} , \quad \widetilde{j}^{a}(w)  j^b(z)\sim {f^{a b c}\  j^c(w) \over z-w} , \quad    \widetilde{j}^a(z)\widetilde{j}^{b}(w)\sim reg \\
&&\bar{j}^a(\bz) j^{b}(w)\sim {f^{a b c}\  j^c(w) \over \bz-\bw},  \quad  \widetilde{j}^b(z)\bar{j}^{a}(\bw) \sim {f^{a b c}\  \bar{j}^c(\bar{w}) \over z-w}\nonumber
\ea
%\ba\label{eq:jshjalg2}
%&&j^a(z) j^{b}(\bw)\sim {f^{a b c}\  j^c(\bw) \over z-w} , \quad  j^a(z)\widetilde{j}^{b}(\bw)\sim {f^{a b c}\  j^c(w) \over z-w} , \quad    \widetilde{j}^a(z)\widetilde{j}^{b}(\bw)\sim reg \\
%&&\bar{j}^a(\bz) j^{b}(w)\sim {f^{a b c}\  j^c(w) \over \bz-\bw},  \quad  \bar{j}^{a}(\bz)\widetilde{j}^b(w) \sim {f^{a b c}\  \tilde{j}^c(w) \over \bz-\bw}\nonumber
%\ea
%\mycolor{red}{
%	Maybe the above equation should be (two OPEs are changed, two OPEs have typo with bars.)
%}
This leads to the surprising conclusion that there is a Kac-Moody algebra of $j^a(z)$ and $\widetilde{j}^a(z)$ which closes only on the $j^a(z)$\footnote{This is plausibly the manifestation of a degeneracy in the algebra.}.
Repeating the previous steps we find
\be\label{eq:Tsj2ope}
T^S(z) \widetilde{j}^a(w)\sim \tilde{C}_2 \left[ {1\over (z-w)^2} j^a(w) + {1\over z -w} \partial j^a(w)\right]
\ee
We see that if we identify $\widetilde{j}^a(z)\equiv j^a(z)$ we have agreement with the conformal properties of the operators as currents with weights $(1,0)$. This is very interesting and consistent with the identification of dimension one states in \cite{Pasterski:2017kqt}. Nevertheless, it implies an one-to-two relation between gauge amplitudes and CCFT correlators since in this way any negative helicity gluon is mapped to a positive one. It is plausible that this discrepancy lies in the detailed analysis of the conformal primary wave functions for the dimension one primary and its shadow. In \cite{donnay_conformally_2019} there is a subtle difference between the two operators due to a dimension one logarithmic operator. So it is more sensible to have the relation $\widetilde{j}^a(z)\simeq j^a(z)$ modulo subleading in the limit $\D\to 1$ logarithmic operators. In that sense the Sugawara energy momentum tensor captures the leading conformal properties of the shadow operators. We leave this interesting question for future work.

	\bibliography{celestial_amp}
	
\end{document}